\documentclass{JFM-FLM_Au}
\usepackage[table]{xcolor}
\usepackage{comment}
\usepackage{bm}
\usepackage{hyperref}

\usepackage{caption}
\usepackage{subcaption}

\DeclareCaptionLabelFormat{custom}{\textsc{#1}~#2.}

\lefttitle{Sumit Lonkar, Singeetham Pranaykumar and Pratikash P. Panda}
\righttitle{Journal of Fluid Mechanics}

\title{Hydrodynamic mechanism and suppression of cavity breathing oscillations in a twin-cavity supersonic combustor}

\author{Sumit Lonkar\aff{1}, Singeetham Pranaykumar\aff{1} \and Pratikash P. Panda\aff{1}}

\affiliation{\aff{1}Department of Aerospace Engineering, Indian Institute of Science, Bengaluru 560012, India}

\corresau{Pratikash P. Panda, \email{pratikashp@iisc.ac.in}}

\begin{document}
\maketitle
\begin{abstract}
Low-frequency cavity breathing oscillations can strongly affect flame stabilization and operability in cavity-stabilized supersonic combustors, yet their hydrodynamic origin and suppression remain insufficiently understood for opposed twin-cavity configurations. We investigate the breathing dynamics of an opposed twin-cavity supersonic combustor under low enthalpy and high enthalpy conditions using synchronized high-speed Schlieren imaging, wall-pressure measurements, numerical simulations, nonlinear phase-space reconstruction, Hilbert-transform analysis and reduced-order modelling. Under low enthalpy conditions, a self-sustained breathing mode is observed, with a dominant frequency of approximately 156 Hz and large phase differences between cavity-floor and ramp pressures. The combined measurements and simulations show that periodic mass exchange between the cavity and freestream drives cyclic cavity depressurization and repressurization, producing a pressure-shear layer feedback loop that sustains the oscillation. A two pressure-state reduced order model, coupled to a shear layer displacement coordinate, reproduces the measured frequency, pressure phase relationship and dominant stability characteristics, demonstrating that both bulk cavity and reattachment/shock-foot pressure states are required to represent the dynamics. The response of shear layer oscillation to mass-addition and heat release and is studies. Upstream injection of an inert gas (nitrogen) produces only transient attenuation: the oscillation recovers as the injected gas convects downstream. In contrast, ethylene injection followed by ignition causes permanent suppression, with the nonlinear attractor collapsing to a stable equilibrium and the dominant mode becoming stable. Heat release reorganizes the cavity pressure field and modifies density, compressibility and pressure-response timescales, thereby weakening the feedback responsible for sustaining the breathing mode.

\end{abstract}



\section{Introduction}
\label{sec:introduction}
Among various classes of air-breathing propulsion systems, the supersonic combustion ramjet (scramjet) offers the highest efficiency in the Mach 6–12 flight regime. Continuous advancements in air-breathing propulsion technologies are redefining space access, with multi-stage propulsion systems being proposed as a promising approach for reusable launch vehicles \cite{liu2025research}. For instance,   \cite{jazra2013design} proposed a reusable launch vehicle concept in which a scramjet-based second stage operates over a wide Mach number range to deliver a payload of approximately 100 kg to low Earth orbit, requiring combustion efficiencies of at least 80\%.

A fundamental challenge in scramjet combustors is the extremely short flow residence time (typically $< 1~\text{ms}$) due to the high flow velocities within the combustor \cite{liu2020review,urzay2018supersonic}. This limited residence time makes fuel–air mixing, reliable ignition, and flame stabilization highly challenging, resulting in reduced combustion efficiency.

To enhance mixing efficiency, various fuel injection strategies have been investigated, including normal sonic injection, angled injection, and three-dimensional injection schemes. Additionally, mixing augmentation techniques such as vortex generators and riblets have been explored to improve fuel-air interaction \cite{sheng2024improving,quan2023experimental}.

To improve residence time and achieve stable flame holding, several flame holding devices such as struts, backward-facing steps, and cavities-have been employed. These devices generate low velocity recirculation zones that enhance residence time, promote mixing, and provide localized ignition sources for sustaining combustion. Among these, cavity-based flameholders have demonstrated superior performance due to their non-intrusive nature, reduced total pressure loss, and lower aerodynamic heating. Consequently, cavity configurations have been widely adopted in several experimental and flight demonstration programs \cite{mathur2004investigation, gruber2008hydrocarbon, jackson2015mach}.

Single-cavity flameholders have been widely employed in supersonic combustors because they generate a stable recirculation zone that enhances flame stabilization and fuel-air mixing \cite{quan2023experimental,venkateswarlu2025recent,tuncer2010cavity}. The aero-thermodynamics of a single-cavity have therefore been extensively investigated, and the dominant mechanisms responsible for cavity oscillations are now well established. Pressure fluctuations originate from the interaction between the separated shear layer spanning the cavity opening and the cavity trailing edge. The impingement of the shear layer generates compression waves that propagate upstream and perturb the separating shear layer at the cavity leading edge, establishing a coupled hydrodynamic-acoustic feedback loop that produces self-sustained oscillations.

\cite{heller1975physical} demonstrated that this feedback mechanism is governed by the periodic entrainment and expulsion of fluid through the cavity opening, causing the cavity to behave like a reciprocating pseudo-piston. Subsequent studies showed that the oscillation characteristics depend strongly on cavity geometry. The dominant mode changes with the cavity aspect ratio (cavity length to depth ratio (L/H) ), while increasing the aft-wall angle weakens the primary recirculation zone and reduces shear layer re-compression. Consequently, inclined aft-wall cavities have become the preferred flameholder geometry in scramjet combustors because they suppress Rossiter-type oscillations by weakening the acoustic feedback mechanism \cite{vikramaditya2009effect,yu2001effect}.

Although the instability mechanisms of single-cavities are well understood, recent investigations by \cite{vishnu2019effect} have shown that they are not governed solely by geometry. Heat transfer and combustion modify the cavity pressure field, alter the recirculation zone, and change the development of the shear layer, thereby influencing the oscillation frequency and the mass exchange between the cavity and the external flow. However, these studies have been restricted almost exclusively to isolated single-cavity configurations.

To improve combustion performance, dual-cavity flameholders have recently been proposed. Experimental studies have reported enhanced ignition characteristics, improved mixing, higher combustion efficiency, and increased thrust compared with conventional single-cavity configurations \cite{zhang2022experimental,collatz2009dual}. \cite{rajesh2023implications} further showed that twin cavities improve fuel-air mixing with only a marginal increase in stagnation pressure loss. Depending on the relative placement of the cavities, dual-cavity flameholders are generally classified into tandem and opposed configurations. While previous investigations have primarily examined their effects on combustion performance and pressure oscillations \cite{tan2026flame,wang2015large}, the fundamental hydrodynamic interactions between the two cavities remain largely unexplored .

Unlike an isolated cavity, an opposed dual-cavity configuration introduces additional flow interactions through the mutual coupling of the two shear layers, recirculation zones, and compression-wave systems \cite{tang2024flow}. These interactions can modify the instability mechanism responsible for shear-layer oscillations and significantly influence the unsteady pressure field within the combustor. Despite their importance, the governing hydrodynamic processes and cavity to cavity coupling mechanisms have not yet been systematically investigated.

The present study addressed this knowledge gap by experimentally investigating the hydrodynamic mechanism underlying low-frequency cavity-breathing oscillations in a long, shallow opposed dual-cavity combustor. The combustor was integrated with a direct-connect high-enthalpy test facility and was investigated under both low- and high-enthalpy inlet conditions. The flow field and, under reacting conditions, the flame structure were characterized using synchronized high-speed Schlieren and $CH^*$ chemiluminescence imaging together with time-resolved wall-pressure measurements. Simultaneous pressure and schlieren measurements established the coupling between the cavity pressure field and the breathing motion of the shear layer. These observations were subsequently used to formulate a two-pressure-state dynamical model that captured the characteristic timescale and pressure dynamics of the experimentally observed oscillation. The governing dynamics were further interpreted using three non-dimensional parameters representing the relative pressure-relaxation timescale, pressure forcing and pressure--shear-layer coupling strength.

The response of the breathing mode to upstream injection was then investigated to distinguish the effects of mass addition from those of heat release. Inert $N_2$ injection under low-enthalpy conditions was used to isolate the influence of mass addition, whereas $C_2H_4$ injection followed by combustion under high-enthalpy conditions was used to examine the influence of heat release. The resulting transient stability characteristics were quantified through phase-space reconstruction and instantaneous growth rates obtained using Hilbert-transform analysis. Finally, the experimentally observed stability transitions were interpreted using an integrated compressible Rayleigh formulation and related to the two-pressure-state model, providing a unified framework for describing the measured pressure dynamics, shear-layer response and stability of the cavity-breathing mode.

The manuscript is organized as follows. Section~\ref{sec:experimental_section} describes the experimental facility, combustor configuration and diagnostic techniques. Section~\ref{sec:mechanisum} characterizes the baseline cavity-breathing mode and establishes its underlying hydrodynamic mechanism using synchronized experiments and time-resolved numerical simulations. The two-pressure-state dynamical model is subsequently presented in Section~\ref{sec:two-pressure-model} to quantify the governing pressure--shear-layer dynamics and their dependence on inlet enthalpy. Section~\ref{sec:control} examines the modification and suppression of the breathing mode through inert mass addition and reacting heat addition. The resulting stability transitions are quantified using phase-space and Hilbert-transform analyses and interpreted using the integrated compressible Rayleigh formulation. The principal findings are summarized in Section~\ref{sec:Conclusion}, while the derivation of the integrated Rayleigh equation is provided in Appendix~\ref{app:rayleigh_derivation}.


\section{Experimental methodology and numerical framework}
\label{sec:experimental_section}
\begin{figure}[hbt!]
\centering
\includegraphics[width=0.80\textwidth]{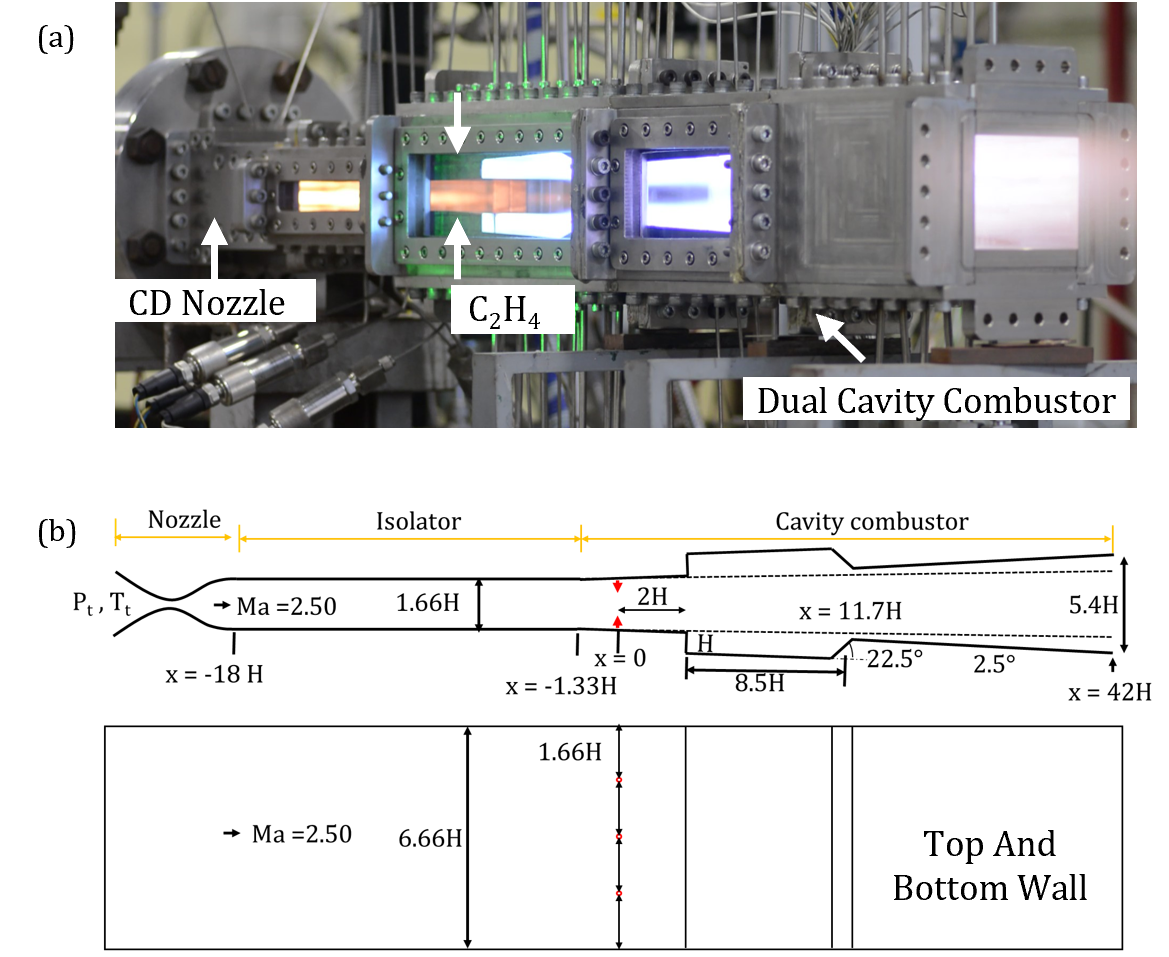}
\caption{(a) Dual-cavity scramjet combustor integrated with the direct-connect high-enthalpy test facility; (b) schematic representation of the experimental configuration.}
\label{mag_1}
\end{figure}

The experiments are conducted in a direct-connect high-enthalpy test facility at the Advanced Propulsion Research Laboratory (APRL) of the Indian Institute of Science, Bangalore, India. Comprehensive details of the facility and its operational characteristics are reported in earlier studies \cite{lonkar2026mode,lonkar2026experimental,thakor2020flame}. For the present study, a dual-cavity scramjet combustor, consisting of a constant area isolator and an opposed dual-cavity diverging combustor section, was mounted at the exit of the Mach 2.5 convergent-divergent (CD) nozzle. The nozzle was connected to an inline methane preheater, as shown in figure~\ref{mag_1}(a). The top and bottom walls of the combustor are geometrically identical and are instrumented with pressure ports along the centerline of the combustor as well as within the cavity at designated locations. The sidewalls of the isolator and cavity sections are fabricated from Corning-7980 fused silica windows, providing optical access for flow and flame visualization.

Figure~\ref{mag_1}(b) presents a schematic of the experimental configuration. All geometric dimensions are normalized by the cavity height, $H$. The origin of the coordinate system is defined at the location of the fuel injector. The isolator height is $1.66H$, and the combustor width is $6.66H$. The isolator length of $16.66H$ was selected to reduce nozzle-combustor interaction and allow sufficient flow development. Although not optimized through a parametric study, the chosen length satisfies the primary objective of isolating the combustor dynamics for the present investigation. Downstream of the isolator, the top and bottom walls diverge at an angle of $2.5^\circ$ relative to the flow direction. Three equally spaced circular injectors of $2$ mm diameter are located on both the top and bottom walls at a distance of $2H$ upstream of the cavity leading edge and serve as a fuel injector ports. The cavity, with an $L/H$ ratio of 8.5, is positioned on the diverging top and bottom walls in an opposed configuration. Cavities with larger $L/H$ ratios, approaching a closed-cavity configuration ($L/H>10$), are generally more favorable for ignition, fuel-air mixing and flame stabilization, particularly for fuels with longer ignition delays; however, such configurations are also susceptible to low-frequency shear-layer oscillations \cite{lonkar2026experimental,wang2026study,bao2015effect}. Accordingly, a cavity with $L/H=8.5$ was selected for the present study, as it provides a configuration in which low-frequency shear-layer oscillations are observed while retaining a relatively open cavity geometry. 

\begin{figure}[hbt!]
\centering
\includegraphics[width=0.80\textwidth]{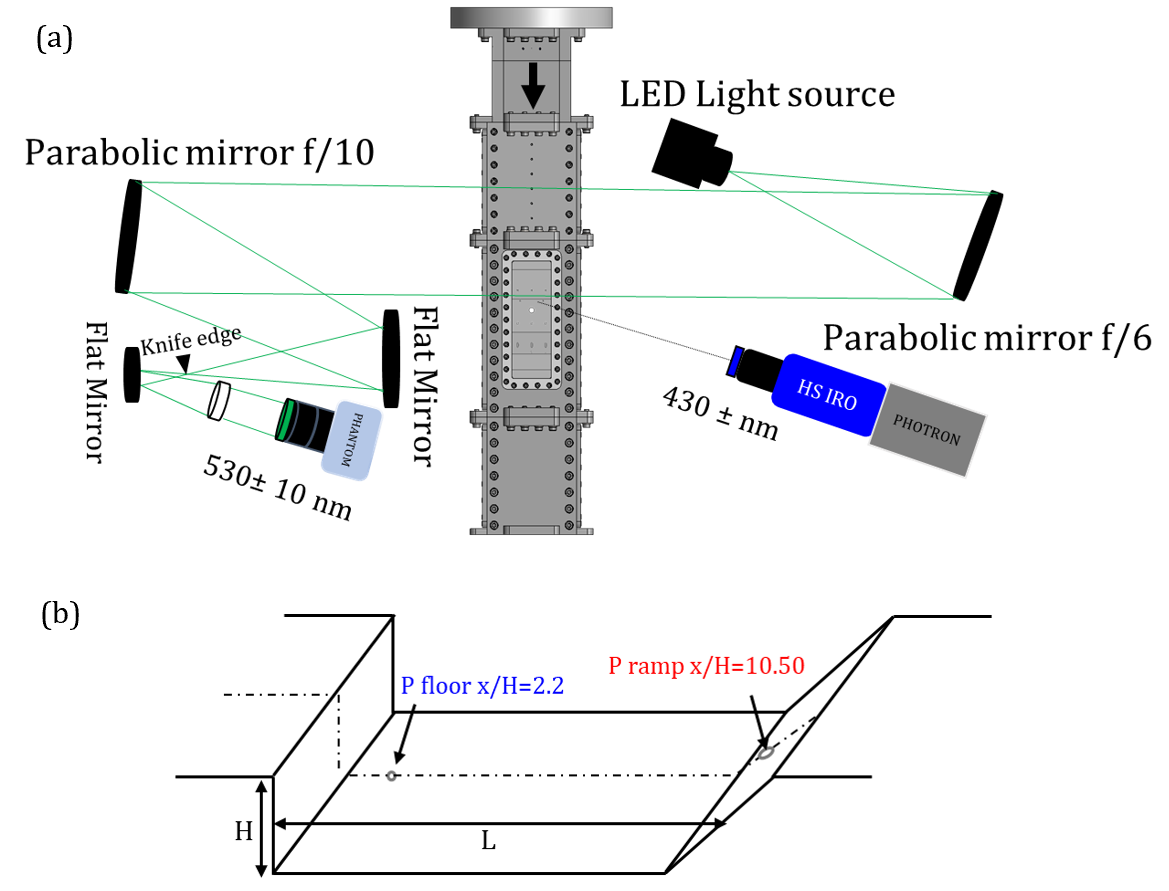}
\caption{ (a) Optical arrangement for synchronized Schlieren and $CH^*$ chemiluminescence diagnostics; (b) geometric schematic indicating the high-frequency absolute pressure transducer locations in the top and bottom cavities.}
\label{mag_2}
\end{figure}

\subsection{Diagnostics}

Simultaneous time-resolved high-speed Schlieren imaging, $CH^*$ chemiluminescence, and high-frequency pressure measurements were performed to characterize the flow field, flame dynamics, and cavity shear-layer oscillations. 

A Z-type Schlieren configuration was employed, as shown in figure~\ref{mag_2}(a). A $290~\mathrm{mm}$ diameter collimated parallel light beam was generated using an $f/6$ parabolic mirror and directed normal to the cavity section. The beam was focused using an $f/10$ collecting mirror onto its focal plane, where a vertically oriented knife edge was positioned to capture density gradients primarily along the streamwise (x) direction. Flat mirrors were used to fold and redirect the optical path to accommodate the available experimental space. Schlieren images were acquired using a Phantom VEO 640 high-speed camera equipped with a $50~\mathrm{mm}$ $f/2.8$ Nikon lens, a 20 mm extension tube, and a $530~\mathrm{nm}$ centre-wavelength (CWL), $\pm10~\mathrm{nm}$ Edmund Optics band-pass filter. Illumination was provided by a Mightex Super High-Power LED Collimator Source, Type-H, with a rated optical power of $60~\mathrm{W}$.

The $CH^*$ chemiluminescence imaging system was positioned at an angle of less than $10^\circ$ relative to the Schlieren optical axis to minimize interference from the Schlieren beam while maintaining the required field of view. Chemiluminescence images were acquired using a Photron FASTCAM SA5 high-speed camera coupled with a LaVision high-speed IRO-X intensifier equipped with a P46 phosphor. The camera was fitted with a $105~\mathrm{mm}$, $f/2.8$ Nikon lens and a $430~\mathrm{nm}$ CWL $\pm10~\mathrm{nm}$ Edmund Optics band-pass filter to selectively capture the $CH^*$ chemiluminescence emission.

Both Schlieren and $CH^*$ chemiluminescence images were acquired simultaneously at a sampling rate of $20~\mathrm{kHz}$, enabling direct temporal correlation between the flow structures and flame dynamics. An exposure time of $1~\mu\mathrm{s}$ was used for the Schlieren imaging, while an exposure time of $8~\mu\mathrm{s}$ and gain value of 45 was used for the $CH^*$ chemiluminescence measurements.

The pressure fluctuations within the cavities were measured using four high-frequency piezoresistive absolute pressure transducers (Endevco-8530C-50PSI, $\pm 0.1$ $\%$) with a sensitivity of $ 4.5 mV/psi$. The raw signal was acquired directly via an NI-9205 programmable input-based c-DAQ module, with an input voltage range of $\pm1$~V, enabling improved sensitivity without a signal amplifier. The raw voltage signal was sampled at $40~\mathrm{kHz}$, which is significantly higher than the frequency of interest, thereby avoiding aliasing effects and reducing uncertainty in the observed dynamics. The raw voltage signal was converted to pressure using the manufacturer’s factory calibration chart and further validated by comparing the absolute pressure reading with static pressure measurements obtained from a WIKA-P30 pressure transducer (0--6bar, $\pm0.006bar$) connected to a common plenum during laboratory calibration. The sensors were mounted along the centerline of both the cavities, with one located on cavity floor near the cavity step at $x/H=2.20$ and another at the mid-ramp region at $x/H=10.50$ as shown in figure~\ref{mag_2}(b). 

Synchronization of all diagnostics was achieved using a Berkeley BNC-577 Pulse generator.

\subsection{Numerical simulations}

The governing equations are discretized using the finite volume method and solved within the OpenFOAM framework by coupling the rhoCentralFoam and reactingFoam solvers, followed a methodology similar to that adopted in previous studies \cite{li2020quasi, marcantoni2017rhocentralrffoam, zhao2016study}. The coupled solver has previously been validated against a few non-reacting and reacting test cases, demonstrating its capability to accurately predict the relevant flow and reacting-flow characteristics \cite{pranaykumar2024insights}. In the present study, the species source term is deactivated to simulate the non-reacting flow field. The rhoCentralFoam solver is widely adopted within the OpenFOAM framework for compressible and supersonic flow simulations.Turbulence closure is achieved using the $k$-$\omega$ SST model.
For spatial discretization, the convective fluxes are evaluated using the Kurganov--Noelle--Petrova (KNP) scheme, a second-order central-upwind method suitable for compressible flows with strong gradients and shock waves. The diffusive terms are discretized using a second-order central-difference scheme. Temporal discretization is performed using a second-order backward differencing scheme. These numerical methods are selected to provide an appropriate balance between numerical stability and accuracy for resolving the shock structures, cavity shear layer, flow field. The governing equations solved using the RANS formulation are described below.

\begin{equation}
\frac{\partial \rho}{\partial t}
+ \nabla \cdot (\rho u) = 0
\tag{1}
\end{equation}

\begin{equation}
\frac{\partial (\rho u)}{\partial t}
+ \nabla \cdot \bigl(u (\rho u)\bigr)
+ \nabla p
+ \nabla \cdot \tau
= 0
\tag{2}
\end{equation}

\begin{equation}
\frac{\partial (\rho E)}{\partial t}
+ \nabla \cdot \bigl(u (\rho E)\bigr)
+ \nabla \cdot (u p)
- \nabla \cdot (\tau \cdot u)
+ \nabla \cdot j
= \dot{\omega}_T
\tag{3}
\end{equation}

\begin{equation}
p = \rho R T
\tag{4}
\end{equation}

Here, $u$ represents the velocity vector, $\rho$ the fluid density, $T$ the temperature, $\tau$ the viscous stress tensor, and $E$ the total energy. The diffusive heat flux is denoted by $j$ and the specific gas constant is $R$. The combustion heat release rate is given by $\dot{\omega}_T$. 

\begin{figure}[!h]
    \centering
    \includegraphics[width=\linewidth]{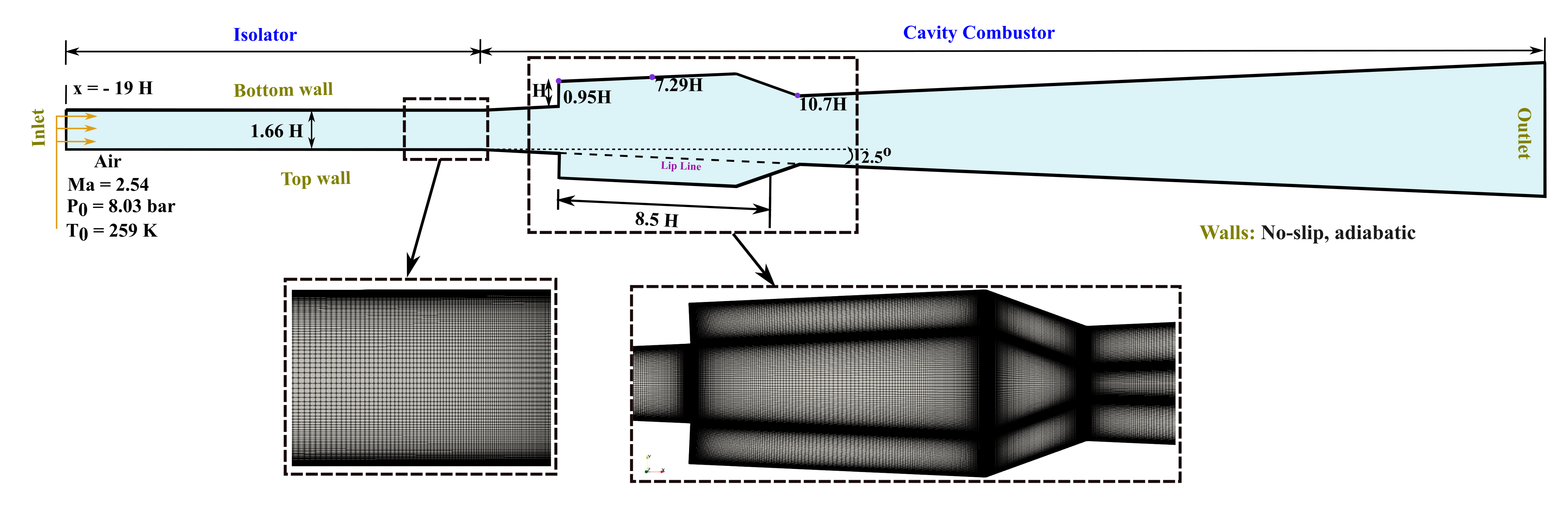}
    \caption{Computational Domain}
    \label{fig:DOMAIN}
\end{figure}

The computational domain consists include isolator and a dual-cavity combustor.The geometric dimensions of the model are illustrated in the figure~\ref{fig:DOMAIN}. A 2-dimensional structured mesh is generated using the blockMesh utility of OpenFOAM and the structure of the mesh at the end of the isolator and cavity section is shown in the figure~\ref{fig:DOMAIN}.

\begin{figure}[!h]
    \centering
    \includegraphics[width=\textwidth]{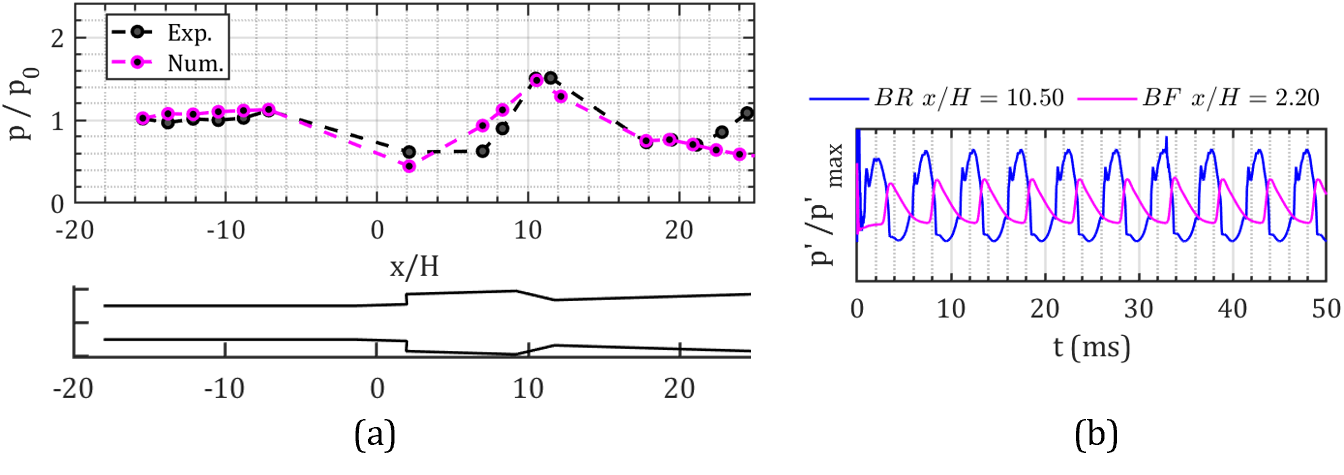}
    \caption{(a) Comparison of wall pressure with experimental data, and (b) Pressure time histories at the ramp and floor location for bottom cavity (CFD)}
    \label{fig:validation2}
\end{figure}

The numerical simulations presented in this study are restricted to the baseline, low enthalpy flow configuration without fuel injection. The objective of these simulations is to characterize the baseline supersonic flow field within the cavity and, in particular, to resolve the unsteady cavity shear-layer dynamics, including shear-layer oscillation, shock-shear-layer interaction, and the resulting recirculation structure. The simulated flow field provides the baseline aerodynamic characteristics required for interpreting the subsequent experimental observations.

Air enters the isolator at a Mach number of $2.54$, with a total pressure of $8.03~\mathrm{bar}$ and a total temperature of $259~\mathrm{K}$. No-slip and adiabatic boundary conditions are imposed on all solid walls, while flow variables at the outlet are extrapolated from the interior of the computational domain. The turbulent kinetic energy and specific dissipation rate are prescribed at the inlet based on the specified turbulence conditions. The computational domain is discretized using a structured grid with adequate near-wall refinement to resolve the boundary layers. The computational mesh consists of approximately $0.33$ million cells.

The predicted bottom-wall pressure distribution was compared with the corresponding experimental measurements for the present low enthalpy, no-injection configuration shown in figure~\ref{fig:validation2}. Along the cavity ramp wall, particularly in the region $8.24 < x/H < 10.69$, deviations exceeding $10\%$ were observed. This discrepancy is primarily attributed to the inherently unsteady nature of the cavity flow, including shear-layer flapping and the interaction between the oscillating shear layer, shock waves, and cavity recirculation region. Furthermore, the numerical pressure measurements at each probe location were averaged over only 10 cycles, which may not fully represent the long-time mean pressure of the highly unsteady flow.

Despite these differences, the numerical predictions capture the overall flow characteristics and pressure trends observed experimentally, including the principal features of the oscillating cavity shear layer and associated shock structures. Since the present CFD analysis is intended primarily to establish the baseline low enthalpy flow field and to provide qualitative and relative comparisons of the cavity flow characteristics, the observed deviations do not affect the principal conclusions drawn from the numerical analysis. The high enthalpy with fuel injection is subsequently investigated experimentally using the fabricated optically accessible combustor.

\section{Phenomenology and physics of baseline shear-layer oscillations}
\label{sec:mechanisum}
The table~\ref{tab:test_conditions} summarizes the inlet conditions and associated measurement uncertainty for the results discussed in the subsequent section.

\begin{table}
\begin{tabular*}{\textwidth}{@{\extracolsep{\fill}}lccccc}
Case & $p_0$ (kPa) & $T_t$ (K) & $Ma$ &
$\dot{m}_{\mathrm{total}}$ (kg/s) & $J$ \\
\hline
Low enthalpy test
& $42.7 \pm 0.64$
& $277 \pm 30$
& $2.54 \pm 0.003$
& $1.55 \pm 0.015$
& -- \\

high-enthalpy test
& $43.5 \pm 0.64$
& $1612 \pm 30$
& $2.47 \pm 0.003$
& $0.58 \pm 0.015$
& -- \\

Low enthalpy test + $\mathrm{N_2}$ injection
& $42.7 \pm 0.64$
& $277 \pm 30$
& $2.54 \pm 0.003$
& $1.55 \pm 0.015$
& $1.85$ \\

high-enthalpy test + $\mathrm{C_2H_4}$ injection
& $43.5 \pm 0.64$
& $1612 \pm 30$
& $2.47 \pm 0.003$
& $0.58 \pm 0.015$
& $1.38$ \\
\end{tabular*}
\caption{Summary of test conditions. All six injectors are choked for
the injection cases. The momentum flux ratio $J$ is defined for an
individual jet.}
\label{tab:test_conditions}
\end{table}

\begin{figure}[hbt!]
\centering
\includegraphics[width=0.95\textwidth]{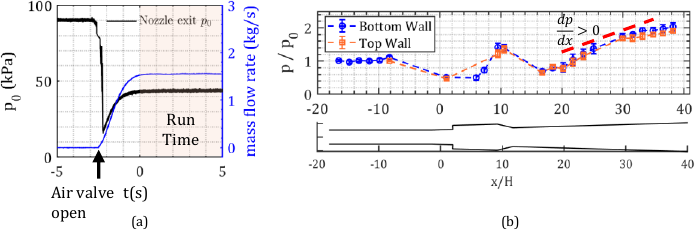}
\caption{(a) Typical facility operation during steady run; (b) normalized time-averaged static pressure distribution along the centerline of top and bottom walls.}
\label{mag_5}
\end{figure}

The shear-layer oscillations in an opposed dual-cavity configuration were first examined under low enthalpy conditions by conducting experiments without igniting the preheater. Room-temperature air was pushed through the test article at a mass flow rate of 1.5 kg/s, establishing an isolator inlet condition of $p_0=42.7 \pm 0.64$ kPa static pressure, Mach $2.54$, and a total temperature of $T_t=277\pm 30$ K. The static pressure was measured using a WIKA-P30 absolute pressure transducer ($6$ $\pm0.006$ bar ) connected to pressure ports on top and bottom walls. The Mach number was estimated from the ratio of the total pressure measured upstream in the preheater to the static pressure measured at the nozzle exit along the top and bottom walls. These measurements were independently validated using a Pitot probe positioned at the nozzle exit during a separate calibration test. Once steady flow conditions were established (defined as time t = $0$ s), as shown in figure~\ref{mag_5}(a), static pressure, high-speed schileren images, and high-frequency pressure data were recorded for $2$ s.

Figure~\ref{mag_5} (b) presents the normalized time-averaged static pressure distribution along the centerline of the top and bottom walls. Here $p_0$ is the absolute static pressure at the exit of nozzle, which corresponds to entry of isolator. A slight pressure rise is observed within the isolator due to viscous effects. This is followed by a sharp decrease( $\approx 50\%$) in pressure across the cavity step region ($2.0 <x/H<8$) due to local flow expansion. A subsequent pressure increase near the cavity ramp indicates shear-layer reattachment accompanied by oblique shock formation ($8.0 <x/H<12.0$). Downstream of the ramp, the pressure decreases progressively in the diverging section up to approximately $x/H=22.0$. Beyond this location, the separation occurs due to the imposed atmospheric back pressure of $90$ kPa, resulting in a pressure rise that matches the ambient exit condition. The separation location is sufficiently downstream of the cavity section; therefore, back pressure does not introduce uncertainty into the cavity dynamics discussed in the present study.

\begin{figure}[hbt!]
\centering
\includegraphics[width=\textwidth]{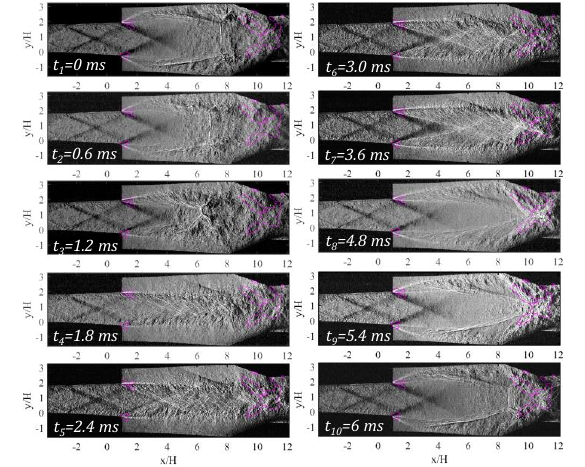}
\caption{Instantaneous schileren images illustrating one complete cycle of shear-layer oscillation. The images are overlaid with contours of maximum standard deviation.}
\label{mag_6}
\end{figure}

The schileren images acquired during steady operation reveal continuous transverse bulk motion of the shear layer in both cavities. The instantaneous schileren images show one complete cycle of shear-layer oscillation in figure~\ref{mag_6}. The reference time, $t_{1}=0$ ms, corresponds to the instant when the shear-layer fully reattaches to the bottom wall of the cavity. The instantaneous schileren images are overlaid with contours of maximum standard deviation computed from images acquired over a $500$ ms time interval. Although the images were initially sampled at a uniform interval of $0.6$ ms, selected frames are presented at non-uniform intervals to better capture the evolution of key features. The flow field appears symmetric about the combustor centerline, consistent with the time-averaged pressure distribution discussed earlier. At $t_{1}=0$ ms, when the shear-layer reattaches to the bottom wall, the reattachment shock terminates at the normal Mach stem. The height and location of the normal Mach stem vary with the instantaneous vertical position of the shear layer ($t_{1}<t<t_{3}$). As the shear layer lifts and becomes nearly parallel to the bottom wall at $t_{4}$, it reattaches outside the cavity; the reattachment shock and normal Mach stem structure disappear. Subsequently, the shear layer remains nearly flat for a brief interval before dipping back into the cavity at $t_{7}=3.6$ ms. As the shear layer penetrates deeper into the cavity, the reattachment location shifts upstream with respect to cavity leading edge, forming two reattachment shocks that interact near the combustor centerline. With further downward motion, the reattachment shocks strengthen and again terminate at a normal Mach stem. The Mach stem grows in size and moves upstream until the shear-layer fully reattaches to the bottom wall, completing one oscillation cycle. This periodic behaviour persists as long as steady inflow conditions are maintained in the combustor. The schileren images further indicate that the downward motion of the shear layer is associated with the formation of expansion waves near the cavity leading edge. The resulting reduction in cavity pressure promotes shear-layer deflection into the cavity, suggesting that the oscillation is driven by a coupled pressure-shear-layer feedback mechanism.

\begin{figure}[hbt!]
\centering
\includegraphics[width=0.80\textwidth]{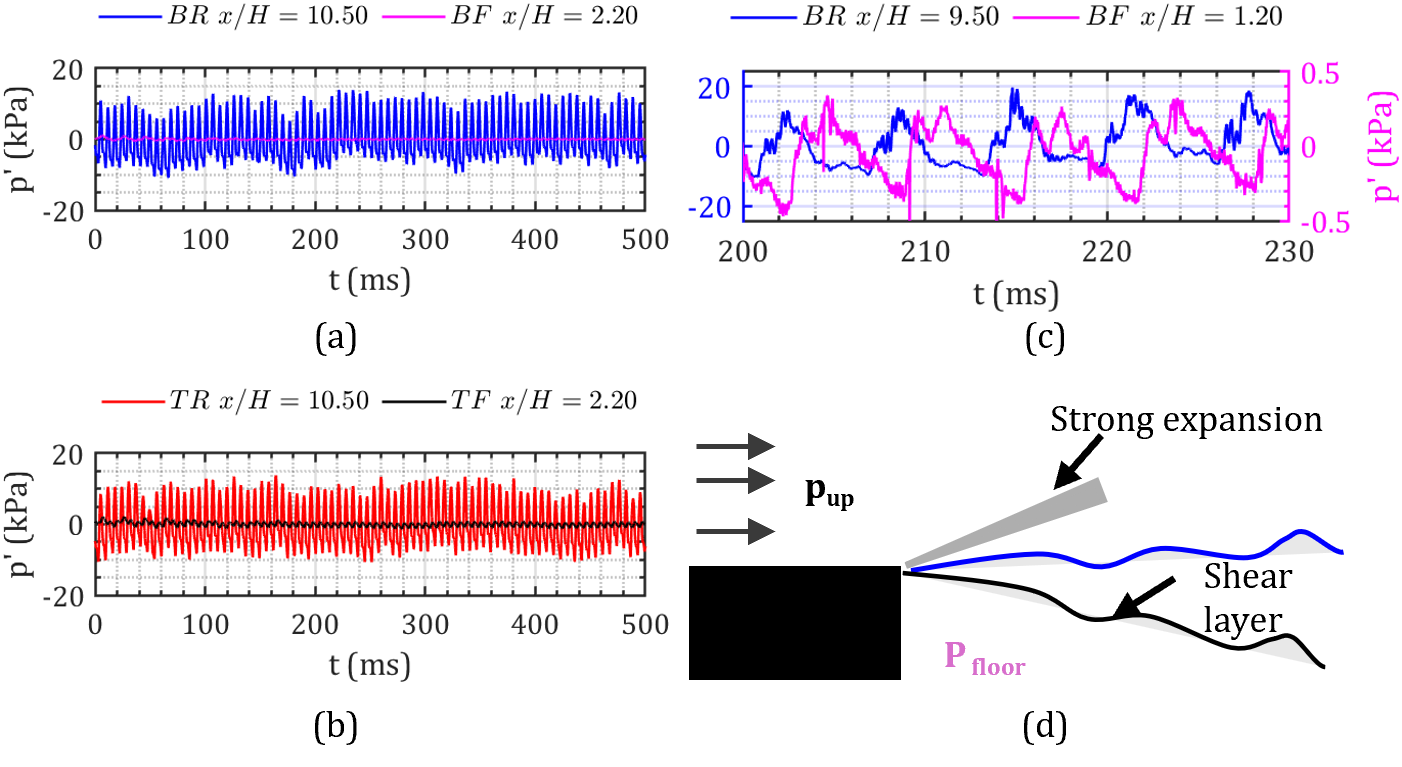}
\caption{ Pressure time histories at the ramp and floor location for (a) the bottom cavity and (b) the top cavity; (c) magnified view of the bottom cavity pressure signal between 200 and 230 ms; (d) conceptual schematic illustrating the pressure-shear-layer coupling mechanism.}
\label{mag_7}
\end{figure}

The dynamic evolution of the shear layer and associated shock structure is closely coupled with pressure fluctuations within the cavity. Therefore, the pressure signals recorded at the cavity ramp and near the cavity step were analyzed, and the corresponding pressure fluctuations magnitudes are presented in figure~\ref{mag_7}(a), (b) for the bottom and top cavities, respectively. For clarity and consistency in subsequent sections, the pressure measurement locations are denoted as BR (bottom ramp), TR (top ramp), BF( bottom floor), and TF (top floor). These abbreviations will be used throughout the remainder of the manuscript. The ramp pressure exhibits a peak-to-peak amplitude of approximately $ 30$ kPa, which is 50 $\%$ of the maximum pressure, and is very detrimental to practical systems; it should be controlled. In contrast, the pressure measured near the cavity step shows significantly lower fluctuation amplitude. The pressure time series reveals a periodic oscillation with nearly constant amplitude, suggesting sustained limit-cycle behaviour. To further illustrate the oscillatory characteristics, a zoomed segment of the signal between 200 and 230 ms is shown in figure~\ref{mag_7}(c). The zoomed-in view highlights a consistent phase difference between the ramp and cavity floor pressure signals. The ramp pressure reaches its maximum when the shear layer reattaches strongly, and a pronounced normal Mach stem forms, while the minimum pressure near the cavity floor coincides with the strong expansion at the cavity leading edge. The ramp pressure also exhibits a brief time period before rising again, indicating a quasi-stationary phase in the oscillation cycle. 

Based on these observations, a schematic representation is developed to illustrate the mechanism shown in figure~\ref{mag_7}(d): the formation of expansion waves reduces the cavity floor pressure $\mathrm{P}_{\mathrm{floor}}$, promoting downward deflection of the shear layer and eventual reattachment to the cavity floor. The temporal behaviour of both pressure signals aligns closely with the flow evolution observed in the schileren images, supporting the presence of a coupled pressure-shear-layer feedback mechanism governing the oscillation dynamics.

\begin{figure}[hbt!]
\centering
\includegraphics[width=0.85\textwidth]{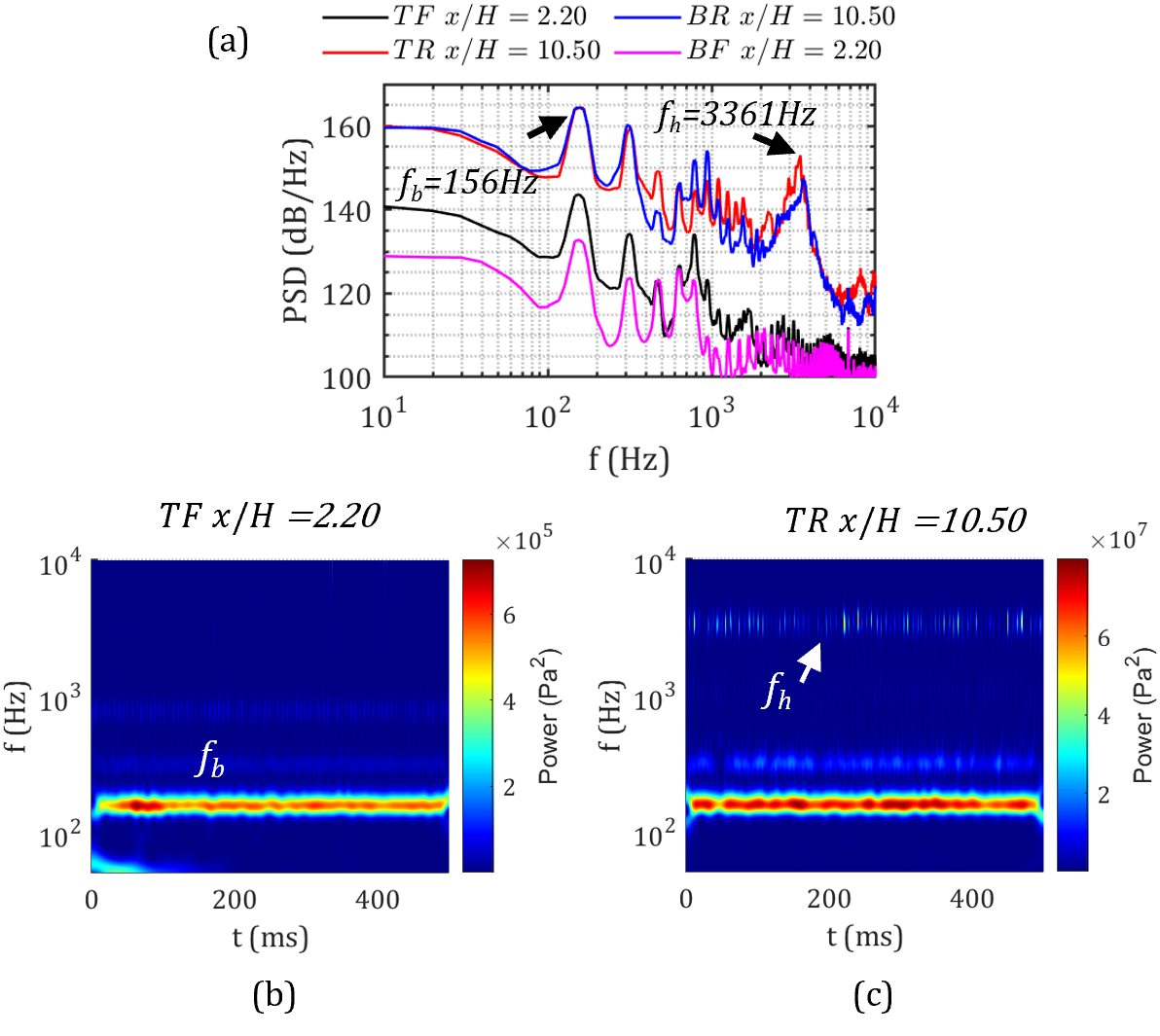}
\caption{(a) Power spectral density of pressure signal in frequency domain; continuous wavelet transform of (b) top cavity floor pressure and (c) top cavity ramp pressure. Here, $f_{b}$ denotes the breathing frequency, and $f_{r}$ represents the cavity Rossiter mode.}
\label{mag_8}
\end{figure}

The pressure signals were analyzed in the frequency domain by computing power spectral density (PSD) using Welch's method. A window length of 4096 samples with 50$\%$ overlap and a Hann window was employed. For a 0.5 s dataset acquired at $40$ kHz, this configuration yields a frequency resolution of $9.77$ Hz and 8 segments. The resulting PSD distribution is shown in figure~\ref{mag_8}(a). All pressure sensors exhibit a dominant frequency at approximately 156 Hz, identified as the shear-layer breathing frequency ($f_{b}$), as it closely matches the period of one cycle observed in the time-domain analysis. The ramp pressure signals (TR and BR) display higher energy content than the cavity floor signals (TF and BF), attributed to multiple shock interactions and stronger reattachment dynamics near the cavity ramp. The second harmonic of the breathing frequency is also prominent in the PSD spectrum, indicating nonlinear oscillatory behaviour. The ramp pressure spectrum further captures the high-frequency mode at approximately $f_{h}= 3361$ Hz. 

The continuous wavelet transform (CWT) of the TF and TR pressure signals is shown in figure~\ref{mag_8}(b) and (c), respectively. The CWT of TF exhibits a strong, persistent signature at $f_{b}=156$ Hz throughout the acquisition period. Similarly, TR exhibits a dominant component at the breathing frequency, while a higher-frequency signature at $f_b = 3623$ Hz appears intermittently. The intermittent occurrence of this component suggests that the associated flow dynamics are not persistent throughout the oscillation cycle. Its temporal occurrence appears to coincide with instances of shear-layer reattachment near the ramp, as observed in the instantaneous Schlieren images. This observation may indicate a possible relationship between the intermittent high-frequency signature and the unsteady evolution of the shear layer. However, the present observation is preliminary, and further analysis of the velocity field is required to establish the underlying flow mechanism and its relationship to the shear-layer reattachment dynamics.

\subsection{Mechanism of Shear-Layer Oscillation}

\begin{figure}[hbt!]
\centering
\includegraphics[width=0.90\textwidth]{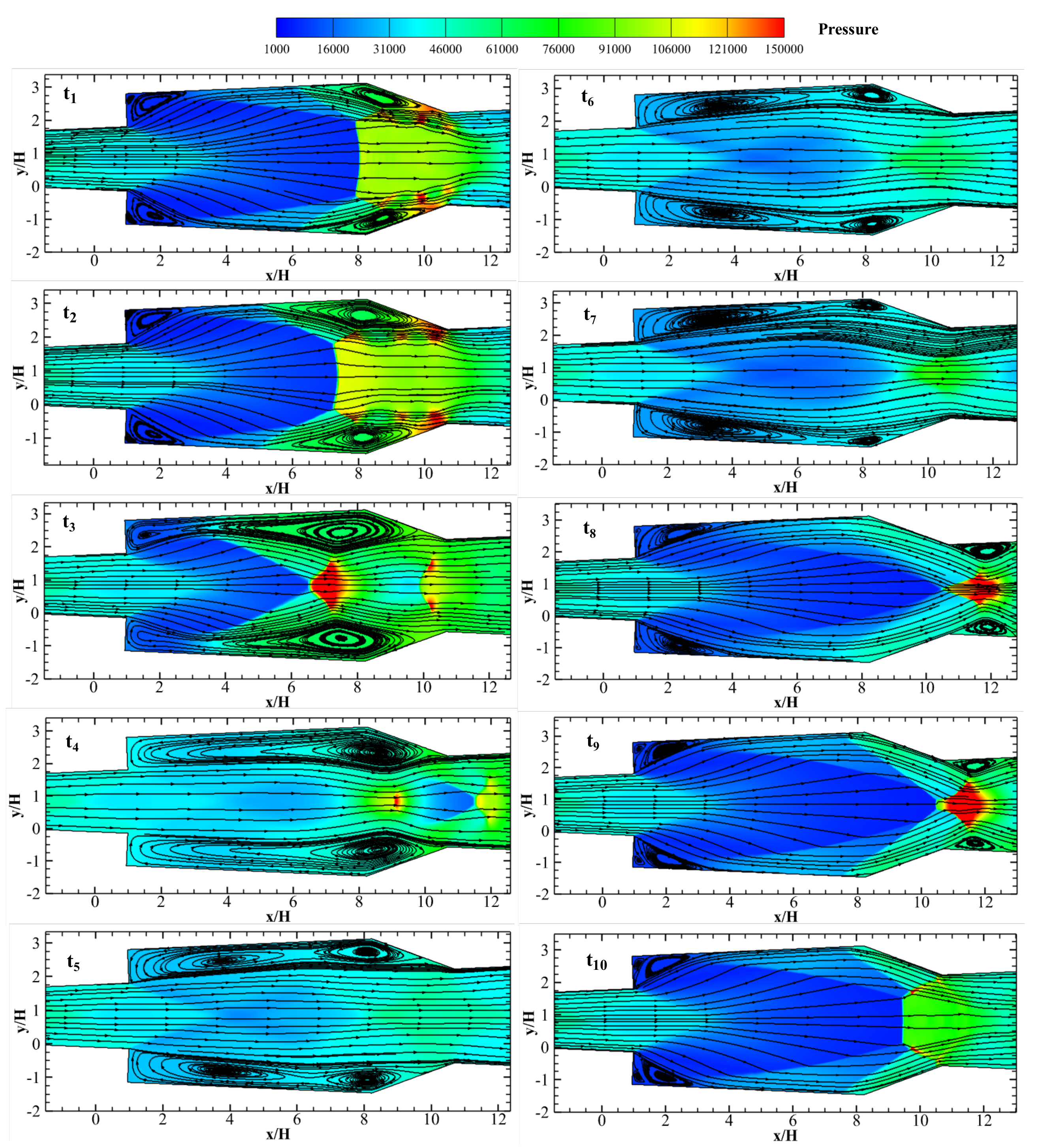}
\caption{Instantaneous contours of pressure overlaid with stream line showing the evolution of recirculation region within the cavity over one cycle of shear layer oscillation.}
\label{mag_9}
\end{figure}
The preceding section established the presence of a coherent low-frequency cavity-breathing mode through time resolved synchronized pressure measurements and Schlieren imaging. To further elucidate the underlying flow physics, the experimentally observed flow evolution is examined using time-resolved numerical simulations.

While the Schlieren images capture the evolution of the shear layer and the associated shock structures, they do not fully resolve the recirculation dynamics within the cavity. Figure~\ref{mag_9} therefore presents instantaneous pressure contours overlaid with velocity streamlines from the low enthalpy numerical simulations at representative phases of the oscillation cycle.

When the shear layer reattaches to the cavity floor ($t=t_1$), a relatively small primary recirculation region is observed beneath the shear layer, accompanied by a secondary vortex in the vicinity of the interaction between the reattachment shock, Mach stem and slip lines. As the shear layer subsequently lifts ($t=t_3$),the primary recirculation region expands substantially, accompanied by an increase in the cavity pressure and a further upward displacement of the shear layer. At ($t=t_4$), the shear layer reattaches downstream of the cavity, following which the cavity pressure decreases and the shear layer moves back towards the cavity floor, initiating the subsequent oscillation cycle. These observations indicate that the low-frequency breathing mode is associated with the coupled evolution of the cavity recirculation, pressure field and shear layer position.

The evolution of the cavity recirculation regions is associated with a continuous exchange of mass between the cavity and the external flow. A reduction in the size of the primary recirculation region corresponds to net mass expulsion from the cavity, resulting in a decrease in cavity pressure and a downward displacement of the shear layer. Conversely, growth of the recirculation region is associated with net mass accumulation within the cavity, leading to an increase in cavity pressure and an upward displacement of the shear layer. 

\begin{figure}[hbt!]
\centering
\includegraphics[width=0.60\textwidth]{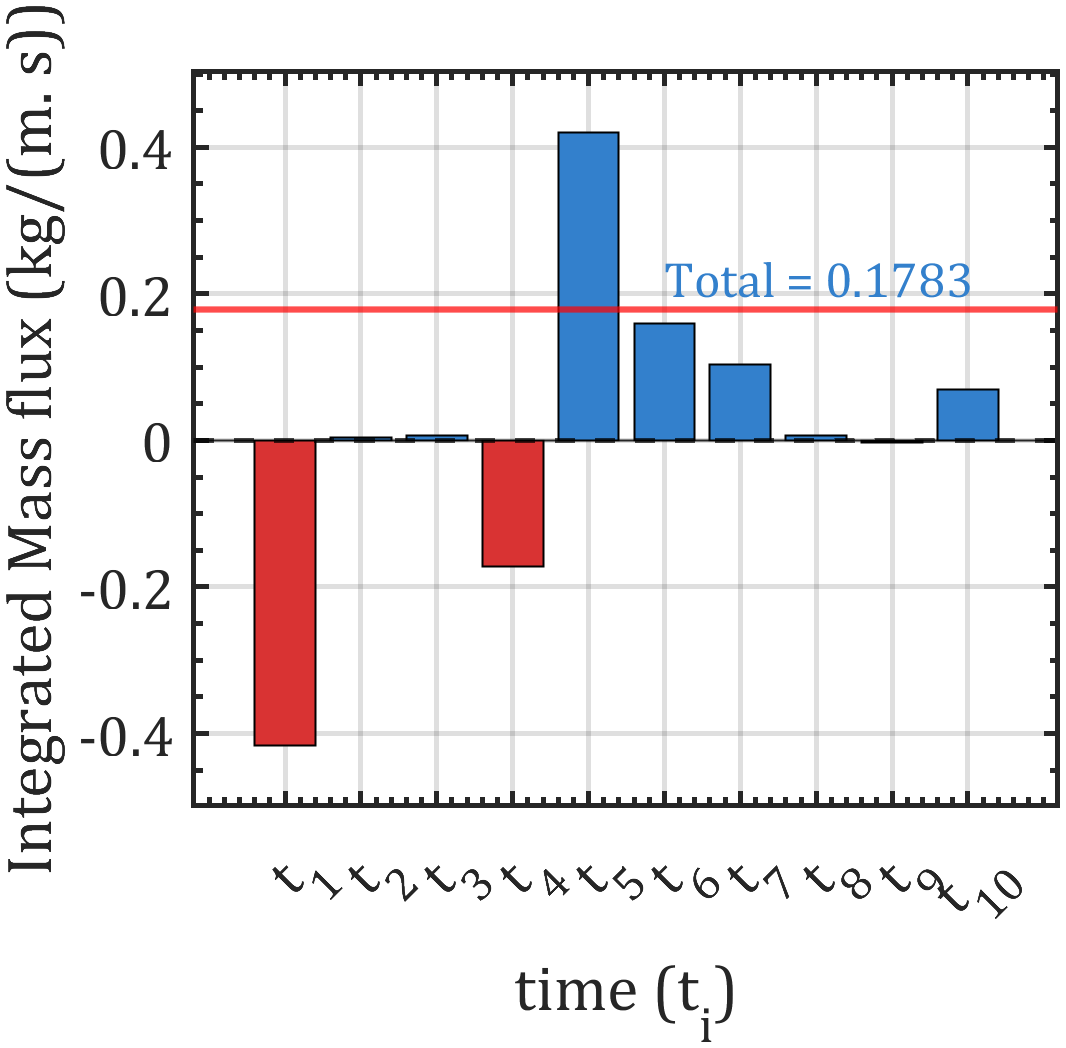}
\caption{ Variation of integrated mass flux calculated along a line connecting the leading-edge to the trailing-edge corner.}
\label{mag_10}
\end{figure}

To quantify this mass exchange, the net mass flux across an imaginary control surface connecting the leading and trailing edges of the cavity is evaluated following the approach reported by \cite{gao2024transition} as:

\begin{equation}
\dot{m}(t_i)''
=
\int
\rho u_n \,dx,
\end{equation}

where $\rho$ is the local density and $u_n$ is the velocity component normal to the control surface. Positive values denote net mass outflow from the cavity, whereas negative values correspond to net mass entrainment into the cavity. The instantaneous integrated mass flux is presented in figure~\ref{mag_10} for each time instant $t_i$. The net mass flux over one oscillation cycle is also indicated by the red line, showing that the cycle-averaged mass flux is directed out of the cavity. This net mass imbalance provides a sustained contribution to the cavity-pressure dynamics and, consequently, to the observed oscillatory behaviour.

\begin{figure}[hbt!]
\centering
\includegraphics[width=0.70\textwidth]{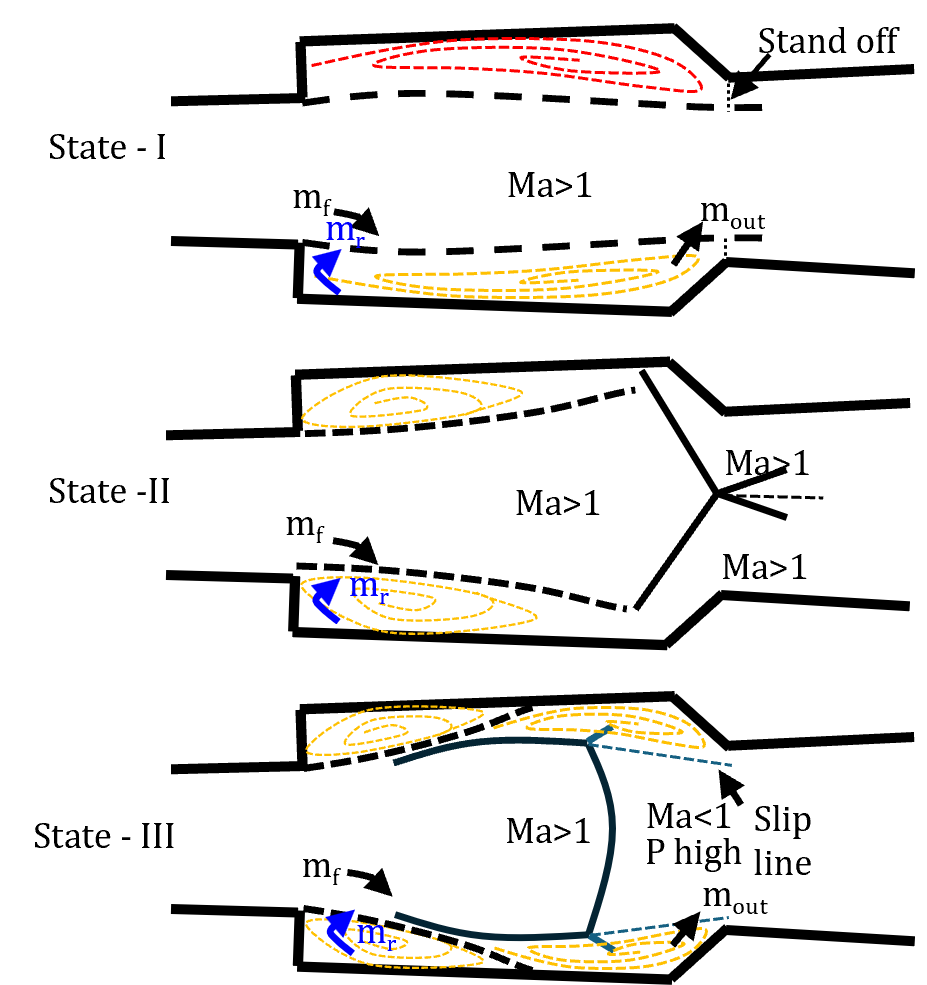}
\caption{Schematic representation of the three characteristic states of the cavity-breathing oscillation}
\label{mag_11}
\end{figure}

The resulting mass-exchange mechanism is summarized schematically in figure~\ref{mag_11}. Three characteristic flow states are identified over one breathing cycle.
\begin{itemize}
\item \textbf{State I:} The shear layer remains nearly parallel to the combustor wall while a large primary vortex occupies the cavity. Net mass is expelled through the cavity ramp, reducing the cavity pressure and initiating downward motion of the shear layer.

\item \textbf{State II:} As the shear layer penetrates into the cavity, freestream entrainment increases while a portion of the flow continues to exit through the reattachment region. The cavity pressure continues to decrease, driving the shear layer towards the cavity floor.

\item \textbf{State III:} Following reattachment, the secondary vortex strengthens and the high-pressure region behind the Mach stem suppresses downstream mass expulsion. The resulting accumulation of mass increases the cavity pressure, forcing the shear layer away from the cavity floor and completing the oscillation cycle.
\end{itemize}

These observations indicate that the cavity-breathing mode is sustained by a pressure-shear layer feedback loop, in which periodic mass exchange between the cavity and the freestream produces cyclic depressurization and re-pressurization of the cavity, accompanied by the corresponding oscillation of the shear layer.

\subsection{Effect of inlet temperature on  cavity-breathing mode}
In the preceding section, the cavity shear-layer oscillation was investigated at a total temperature of $T_t=277\pm30$ K. Although this condition does not represent flight-relevant high-enthalpy conditions, it provides a well-controlled baseline for examining the underlying flow physics. To investigate the evolution of the shear-layer oscillation under high-enthalpy inlet conditions, experiments were subsequently conducted at a total temperature of $T_t=1612\pm30$ K, a static pressure of $p_0=43.5\pm0.64$ kPa, and a Mach number of $M=2.47\pm0.003$. The heated flow was generated using a methane-based preheater.

\begin{figure}[hbt!]
\centering
\includegraphics[width=0.90\textwidth]{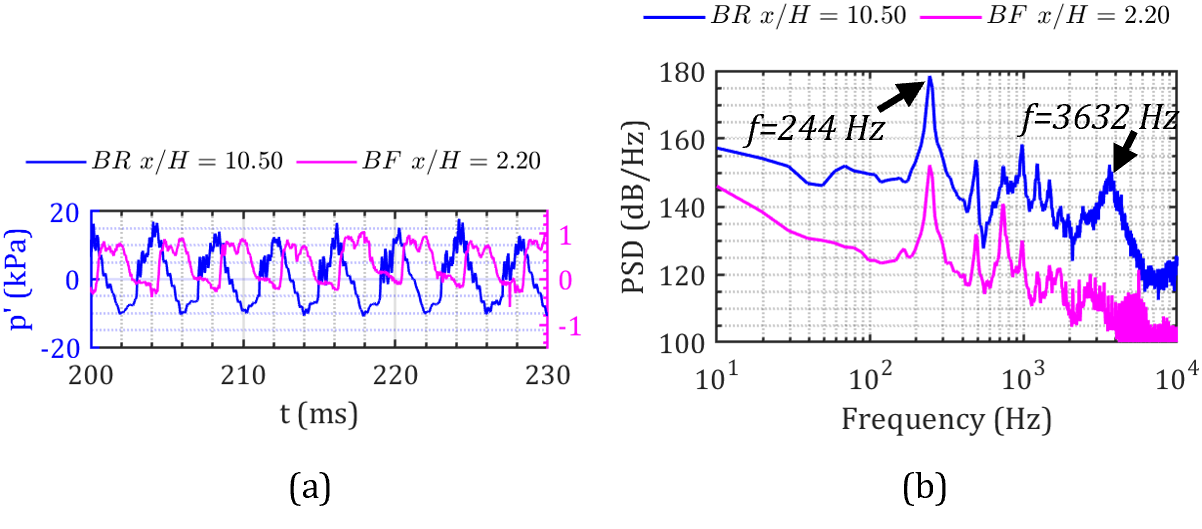}
\caption{(a) Magnified view of the cavity-floor pressure signal between 200 and 230 ms under heated-inlet conditions; and (b) corresponding power spectral density in the frequency domain.}
\label{mag_12}
\end{figure}

The cavity-breathing mode persists under the heated-flow condition; however, both its amplitude and characteristic frequency increase substantially. Figures~\ref{mag_12}(a) and (b) present representative pressure signals measured at the bottom cavity floor and the corresponding power spectral densities, respectively. The pressure signatures measured along the top cavity exhibit similar behaviour to that observed under the low-enthalpy condition and are therefore not shown here. Compared with the low enthalpy condition, the amplitude of the pressure fluctuations increases by approximately a factor of two, indicating a stronger unsteady response of the coupled cavity shear layer and shock system. The dominant breathing frequency increases from approximately $156$ Hz under low enthalpy conditions to $244$ Hz under heated-flow conditions, accompanied by a substantial increase in the spectral peak magnitude. Additional higher-frequency peaks observed in the ramp-pressure spectrum.

The increase in breathing frequency should not be interpreted as a direct consequence of combustion, since the measurements correspond to a high enthalpy heated flow. Instead, the increase in frequency is primarily associated with the higher characteristic velocity scales resulting from the elevated total temperature, including increased convective velocities and acoustic wave speeds. The simultaneous increase in oscillation amplitude indicates that the pressure-shear layer coupling remains strong despite the reduced characteristic timescale.

\begin{figure}[!h]
\centering
\includegraphics[width=0.70\textwidth]{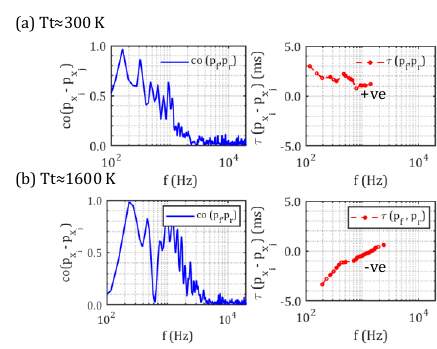}
\caption{Squared magnitude of coherence, $CO(p_f,p_r)$, and corresponding time delay between the bottom cavity-floor pressure, $p_f$, and ramp pressure, $p_r$, for (a) low-enthalpy inlet conditions, $T_t\approx300~\mathrm{K}$, and (b) high-enthalpy inlet conditions, $T_t\approx1600~\mathrm{K}$.}
\label{mag_13}
\end{figure}

To further quantify the coupling between the cavity-floor and ramp pressures, cross-spectral coherence and narrow-band time-delay analyses were performed using the simultaneously acquired pressure signals. Figure~\ref{mag_13}(a) and (b) compares the coherence and corresponding time delay for the low and high enthalpy flow conditions respectively. In both cases, the dominant breathing mode exhibits coherence approaching unity, indicating that the two pressure signals are strongly correlated and form part of the same global oscillatory response. However, the phase relationship changes substantially with increasing inlet temperature.Under low enthalpy conditions, the measured time delay indicates that the pressure fluctuation at the cavity floor precedes that at the ramp. In the high enthalpy flow condition, the sign of the measured time delay reverses, accompanied by a substantial reduction in its magnitude. 

These observations indicate that increasing the total temperature modifies the temporal coupling between the cavity-floor and ramp pressure fluctuations and substantially reduces the characteristic response time. This temperature-dependent change in pressure coupling provides further evidence of the modified cavity dynamics and is examined quantitatively using the reduced-order model presented in the following section.

\section{A Low-order two-pressure-state dynamical model}
\label{sec:two-pressure-model}

The cavity-breathing mode arises from the coupled dynamics of the shear-layer and the cavity pressure field, mediated by periodic mass exchange between the cavity and the external flow. 
Although the breathing mode is governed by a single global oscillation, the wall-pressure measurements reveal that different regions of the cavity respond differently to the underlying flow physics. In particular, the cavity floor ($x/H=2.2$) and cavity ramp ($x/H=10.5$) pressure  signals exhibit coherence values close to unity at the breathing frequency, while maintaining phase differences of approximately $135^\circ$ and $124^\circ$ for the low and high enthalpy flow cases, respectively.

The cavity floor pressure primarily reflects the quasi-uniform pressurization and depressurization of the recirculation region, whereas the ramp pressure is additionally influenced by the unsteady reattachment process and the motion of the shock foot. Consequently, the cavity pressure field cannot be adequately represented by a single pressure variable; instead, at least two dynamically coupled pressure states are required to capture the observed phase lag and spatial evolution of the breathing mode.

Motivated by these observations, a reduced-order linear dynamical model is formulated in which the cavity breathing motion is coupled to two pressure states representing the dominant pressure dynamics within the cavity and at the cavity ramp. The model is intentionally kept minimal so that its parameters can be identified directly from the experimentally measured time-series using least-squares regression, while simultaneously retaining the ability to reproduce the measured breathing frequency, the observed phase relationship between the wall-pressure signals, and the underlying modal stability through eigenvalue analysis.

We assume that, in the narrow band around the breathing frequency, the shear-layer vertical motion can be represented by a single generalized coordinate $\eta(t)$ (e.g., shock/shear-layer displacement), and that wall pressures can be represented by two lumped pressure-like states: cavity floor pressure ($p_f(t)$ ) and cavity ramp pressure ($p_r(t)$).

The model is intended for the band-limited breathing dynamics; it does not resolve Kelvin-Helmholtz roll-up, broadband turbulence, or nonlinear saturation mechanisms.
A two-pressure-state formulation allows $p_f(t)$ to act as a leaky cavity pressure state that depends on the shear layer position, $\eta(t)$ as well as the $p_r(t)$, while $p_r(t)$ acts as a lagged/reactive state driven by both $p_f(t)$ and shear-layer motion, $\eta(t)$. This is the minimal extension that can reproduce the phase shift using real-valued linear dynamics.

We represent the shear-layer breathing coordinate $\eta(t)$ as a second-order oscillator forced by the two pressure states:
\begin{equation}
\ddot{\eta} + 2\zeta\omega_0 \dot{\eta} + \omega_0^2 \eta
=
\alpha_f\, p_f + \alpha_r\, p_r,
\label{eq:eta}
\end{equation}
where $\omega_0$ is the natural breathing frequency of the generalized coordinate, and $\zeta$ is an effective damping ratio. The coefficients $\alpha_f,\alpha_r$ quantify how floor-like and ramp-like pressure loadings accelerate the shear-layer coordinate.

\subsection{Two-pressure-state dynamics}

The floor pressure proxy $p_f$ is modeled as a first-order state driven by the shear-layer displacement (mass exchange/entrainment effect) and relaxing via effective leakage:
\begin{equation}
\dot{p}_f = \beta_{\eta f}\,\eta + a_{f}\,p_f +b_{f}\,p_r,
\label{eq:pf}
\end{equation}
Here $\beta_{\eta f}$ quantifies how $\eta$ drives pressurization/depressurization of the cavity volume, while $a_{f}$ and $b_{f}$ represents the effect of $p_f(t)$ and $p_r(t)$ on the variation of $p_f(t)$.

The ramp pressure $p_r$ is modeled similarly with an explicit self-dynamics and coupling to $p_f$:
\begin{equation}
\dot{p}_r = \beta_{\eta r}\,\eta + b_r\,p_f + a_r\,p_r,
\qquad a_r\le 0.
\label{eq:pr}
\end{equation}
This form allows $p_r$ to behave as a dynamic ``impedance-like'' response to $p_f$ (and to $\eta$), producing a phase shift.

\subsection{State-space form and eigenvalue problem}

Define the state vector
\[
\bm{x} = \begin{bmatrix}\eta & \dot{\eta} & p_f & p_r\end{bmatrix}^T.
\]
Equations \eqref{eq:eta}--\eqref{eq:pr} can be written as $\dot{\bm{x}}=\bm{A}\bm{x}$ with
\begin{equation}
\bm{A}=
\begin{bmatrix}
0 & 1 & 0 & 0 \\
-\omega_0^2 & -2\zeta\omega_0 & \alpha_v & \alpha_r \\
\beta_{\eta v} & 0 & b_f & a_f \\
\beta_{\eta r} & 0 & b_r & a_r
\end{bmatrix}.
\label{eq:A}
\end{equation}
Linear stability and modal content follow from the eigenvalue problem
\begin{equation}
\bm{A}\bm{v}_k = \lambda_k \bm{v}_k,
\end{equation}
where $\lambda_k=\sigma_k+i\omega_k$. The real part $\sigma_k$ is the growth/decay rate and $\omega_k$ is the angular frequency. A complex-conjugate pair with $\sigma>0$ indicates a Hopf-type linear instability (small-signal growth of an oscillation), while $\sigma<0$ indicates a damped mode.

\subsection{Parameter identification from data}

The shear-layer breathing coordinate, $\eta(t)$, is evaluated as the temporal coefficient of the dominant spatial mode obtained from time-resolved Schlieren images using a stationary wavelet transform (SWT) based proper orthogonal decomposition (POD) framework. The Schlieren images represent the spatial distribution of light intensity, $Q(x,y,t)$, which corresponds to the density gradient primarily along the axial ($x$) direction.

The image data are first converted into a space-time matrix, $Q(K,t)$, where $K$ denotes the spatial pixel index. A stationary wavelet transform is then applied to this matrix, and the signal is decomposed into multiple levels using a Daubechies-6 (db6) wavelet, yielding $Q_{\text{level}}(K,t)$. Each decomposition level corresponds to a specific frequency band of the form
\begin{equation}
\left( \frac{f_s}{2^{i+1}}, \; \frac{f_s}{2^i} \right),
\end{equation}
where $f_s$ is the sampling frequency and $i$ denotes the decomposition level.

\begin{figure}[!h]
\centering
\includegraphics[width=0.80\textwidth]{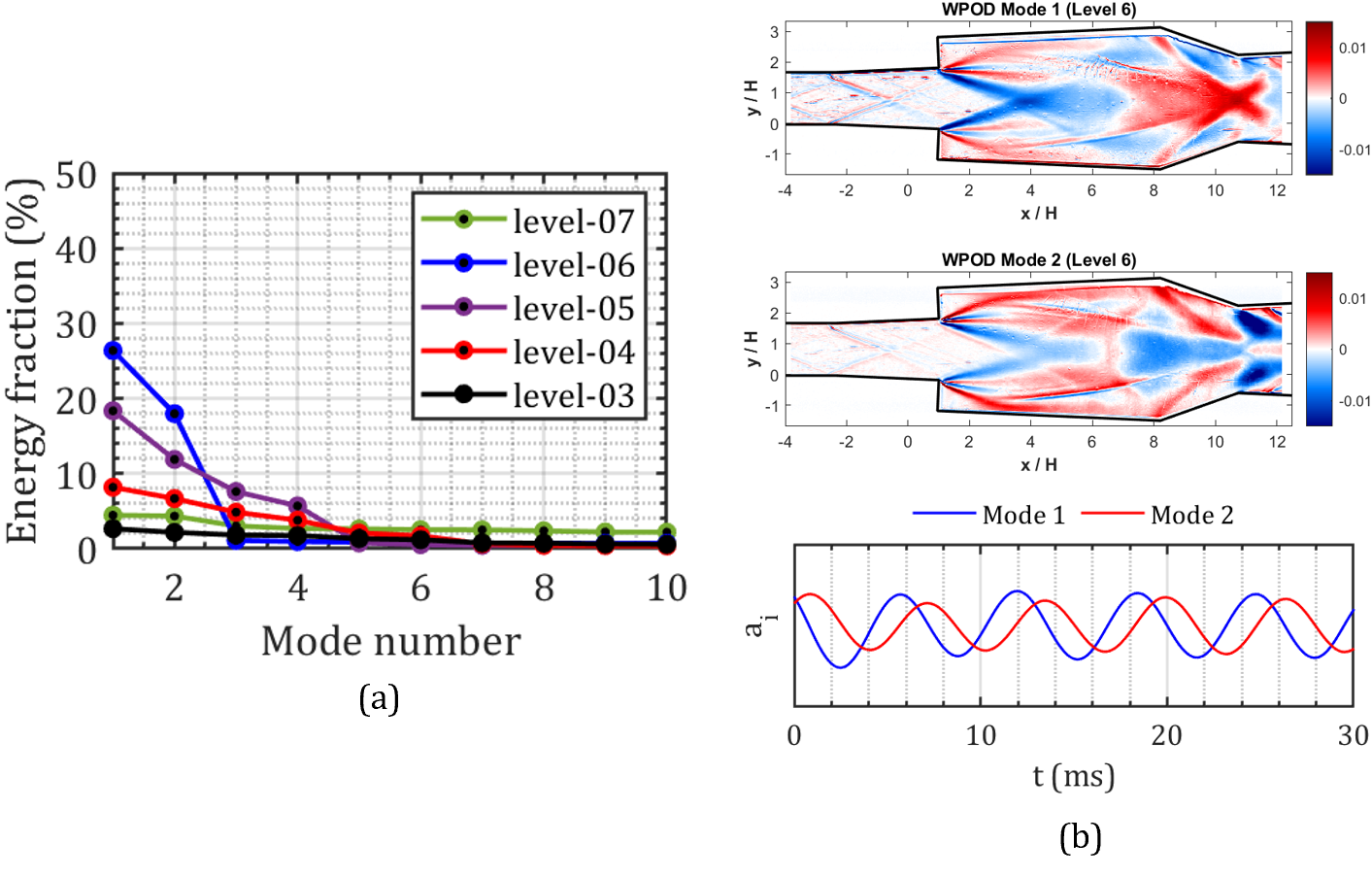}
\caption{(a) Proper Orthogonal Decomposition (POD) energy distribution across different decomposition levels; (b) spatial distribution of the first and second POD modes; and (c) corresponding temporal coefficients.}
\label{fig:swtpod}
\end{figure}

Subsequently, Proper Orthogonal Decomposition (POD) is performed on each filtered level to isolate the dominant coherent structures within the frequency bands of interest identified from the experiments. The resulting spatial POD modes and their corresponding temporal coefficients are shown in figure~\ref{fig:swtpod}.

Among the seven decomposed levels, level 6 contributes the most to the POD energy distribution as presented in figure~\ref{fig:swtpod}. This level corresponds to the frequency band of approximately $156.25$–$315.2~\text{Hz}$, capturing the dominant shear-layer breathing mode and its second harmonic observed in the pressure spectrum. Therefore, the POD mode and temporal coefficient associated with level 6 are selected to define the shear-layer breathing coordinate, $\eta(t)$.           

The temporal coefficient associated with the second dominant mode is used as a measure of the shear-layer breathing coordinate, $\eta(t)$.            

The identified coefficients may be interpreted as the dynamic gains governing the interaction between the shear-layer oscillator and the cavity pressure field. The parameters $\alpha_f$ and $\alpha_r$ quantify the forcing exerted by the cavity floor and ramp pressure states on the shear-layer displacement, while $\beta_{\eta f}$ and $\beta_{\eta r}$ represent the reciprocal forcing of the pressure field by the shear-layer motion. The coefficients $a_f$ and $a_r$ define the intrinsic relaxation rates of the respective pressure states, whereas $b_f$ and $b_r$ describe the dynamic exchange between the floor and ramp pressure fields. Together, these coefficients determine the relative importance of aerodynamic forcing, pressure relaxation and inter-state coupling in sustaining the cavity breathing mode.

For the low enthalpy case, the identified natural frequency,
\[
\omega_0 = 992~\mathrm{rad\,s^{-1}}
\]
corresponds closely to the experimentally measured breathing frequency of approximately $158~\mathrm{Hz}$, confirming that the reduced-order model accurately captures the dominant hydrodynamic timescale. The identified damping ratio ($\zeta=-0.031$) is close to neutral, indicating that the isolated shear-layer oscillator is only weakly stable and that the observed self-sustained oscillation is primarily maintained through pressure feedback from the cavity.

The forcing coefficients reveal a pronounced asymmetry between the two pressure states. The magnitude of the ramp-pressure forcing coefficient,
\[
|\alpha_r| \gg |\alpha_f|,
\]
indicates that fluctuations associated with the reattachment region and shock-foot motion exert a substantially stronger influence on the shear-layer displacement than the bulk cavity pressure. The opposite signs of $\alpha_f$ and $\alpha_r$ further suggest that the two pressure states act in opposing directions on the shear layer, reflecting the competition between cavity pressurization, which tends to lift the shear layer, and the downstream pressure loading associated with shock-foot motion, which promotes downward deflection and reattachment.

The identified pressure relaxation coefficients provide characteristic response times for the cavity pressure dynamics. The ramp-pressure state possesses a finite relaxation time,
\[
\tau_r=\frac{1}{|a_r|}\approx6.9~\mathrm{ms},
\]
which is of the same order as the breathing period
\[
T_b=\frac{1}{f_b}\approx6.3~\mathrm{ms}.
\]
This close correspondence demonstrates that the ramp pressure evolves on essentially the same timescale as the global shear-layer oscillation, implying that the shock-foot and reattachment dynamics actively participate in the feedback loop rather than responding quasi-steadily to the shear-layer motion.

In contrast, the floor-pressure relaxation coefficient is nearly zero, implying an effectively infinite relaxation time for the cavity floor pressure. Rather than behaving as an independent first-order compliance, the floor pressure evolves primarily through forcing from the shear layer and dynamic coupling with the ramp-pressure state. Physically, this indicates that the cavity floor pressure reflects the global cavity pressurization produced by the breathing motion, whereas the ramp pressure behaves as the dynamically responsive component of the pressure field.

The cross-coupling coefficients further demonstrate that the two pressure states form a coupled pressure subsystem. In particular, the large positive value of $b_f$ indicates efficient transmission of pressure disturbances from the ramp towards the cavity interior, while the smaller negative value of $b_r$ represents the feedback from the bulk cavity pressure towards the downstream reattachment region. These coupled interactions generate the experimentally observed phase lag of approximately $135^\circ$ between the two wall-pressure measurements.

Eigenvalue analysis further confirms this physical interpretation. The dominant complex-conjugate eigen pair,
\[
\lambda_{1,2}=-0.545\pm i986.9
\]
represents the global cavity breathing mode, whose small negative real part places the system close to marginal stability. Consequently, relatively weak hydrodynamic forcing is sufficient to sustain the observed limit-cycle oscillation. The second eigen pair,
\[
\lambda_{3,4}=-40.9\pm i361.7
\]
is considerably more damped and corresponds to the internal dynamics of the coupled pressure subsystem. The presence of two oscillatory eigen pairs therefore indicates that the cavity pressure field possesses its own intrinsic dynamics, rather than acting as an algebraic forcing term on the shear layer.

A comparison of the identified symmetric two-pressure-state models for the low ($T_t \approx 300$ K) and high enthalpy ($T_t \approx 1600$ K) cases reveals systematic changes in the breathing frequency, pressure-state dynamics, and coupling structure, while preserving the same dominant global oscillation mechanism.

The dominant breathing frequency increases from approximately $156$ Hz in the low enthalpy case case to about $243$--$247$ Hz in the high enthalpy case. The identified oscillator frequencies $\omega_0/2\pi$ and the eigen-frequencies from the linearized system closely match the spectral peaks in both cases, indicating that the model captures the correct global timescale. The increase in frequency reflects a faster feedback loop in the high enthalpy flow, consistent with the dependence of the characteristic wave speed on $\sqrt{\gamma R T}$ and the modified thermodynamic properties of combustion products relative to air.

Despite this shift in frequency, both cases are governed by a single dominant oscillatory mode. The modal energy analysis shows that essentially all the energy is contained in the leading complex-conjugate eigen pair, while the secondary pair remains strongly damped and dynamically inactive. The dominant eigenvalues in both cases exhibit small negative real parts, indicating weakly damped oscillations close to marginal stability. The high enthalpy case shows slightly stronger damping than the low enthalpy case, suggesting that increased temperature shifts the system marginally away from linear onset while retaining high receptivity. The pressure signals remain highly coherent in both cases, with coherence values close to unity. However, the phase lag between ramp pressure and floor pressure is large ($\sim -130^\circ$) in both cases, indicating that the ramp pressure is not a quasi-steady proxy for cavity pressure but rather a strongly lagged dynamic response associated with reattachment and shock-foot motion. 

The identified parameters show that the ramp-pressure channel dominates the forcing of the shear layer in both cases, as indicated by the larger magnitude of $\alpha_r$ compared to $\alpha_f$. The opposite signs of these coefficients imply that bulk cavity pressure and ramp/shock loading act in competing directions on the shear layer. The pressure–shear coupling coefficients further indicate that the two pressure states respond differently to shear-layer displacement, reinforcing the interpretation of a spatially structured, phase-shifted pressure field. A key distinction between the two cases lies in the ramp-pressure dynamics. In the high enthalpy case, the magnitude of the ramp-pressure relaxation coefficient $|a_r|$ is significantly larger, corresponding to a much faster response of the reattachment/shock-foot region. This leads to a more rapid adjustment of ramp pressure and contributes to the observed increase in breathing frequency. The floor-pressure relaxation coefficient $a_f$ remains near zero in both cases, indicating that the floor-pressure state does not emerge as an independently relaxing compliance variable, but is instead strongly coupled to the ramp-pressure dynamics and shear-layer motion.

Overall, the results indicate that increasing total temperature does not alter the fundamental nature of the cavity breathing mechanism, which remains a single coherent shear–pressure coupled oscillator. However, it significantly modifies the characteristic timescales and strengthens the dynamic role of the ramp-pressure/shock-foot response, leading to higher oscillation frequencies and a more rapidly evolving pressure subsystem.

Table~\ref{tab:cold_hot_params} compares the identified model parameters together with physically meaningful timescales extracted from the regression coefficients. The cavity floor pressure exhibits a fundamentally different behaviour. For both operating conditions, \[
a_f\approx0,\] indicating that no intrinsic relaxation timescale can be identified for the floor-pressure state. Physically, this suggests that the cavity floor pressure behaves as an integral measure of the global cavity pressurization rather than an independently relaxing compliance. Instead, its evolution is governed primarily by forcing from the shear-layer displacement ($\beta_{\eta f}$) and by dynamic coupling with the ramp-pressure state through the coefficient $b_f$.

The forcing coefficients further clarify the physical role of the two pressure states. In both operating conditions, \[|\alpha_r|>|\alpha_f|,\] demonstrating that pressure fluctuations associated with the reattachment region and shock-foot dynamics exert the dominant forcing on the shear-layer oscillator. The opposite signs of $\alpha_f$ and $\alpha_r$ indicate that bulk cavity pressurization and downstream shock loading influence the shear layer in competing directions. The cavity pressure tends to displace the shear layer away from the cavity through volumetric pressurization, whereas the ramp-pressure fluctuations associated with the moving reattachment shock act to restore the shear layer towards the cavity, thereby closing the feedback loop responsible for the breathing oscillation.

The reciprocal coupling coefficients ($\beta_{\eta f}$ and $\beta_{\eta r}$) indicate that shear-layer displacement generates distinct responses in the two pressure states. The stronger response of the ramp-pressure state reflects its sensitivity to relatively small changes in the reattachment location and shock-foot motion, whereas the floor pressure evolves more gradually as the cavity undergoes bulk filling and venting during each breathing cycle.

\begin{table*}
\centering
\renewcommand{\arraystretch}{1.05}
\setlength{\tabcolsep}{7pt}

\begin{tabular}{lccc}
\toprule
\hline
Quantity & $T_t \approx 300$ K & $T_t \approx 1600$ K & Physical interpretation \\
\hline
$f_b$ (Hz)
    & 156 & 246
    & Dominant breathing frequency \\

$T_b=1/f_b$ (ms)
    & 6.41 & 4.07
    & Global feedback period \\

$\omega_0$ (rad s$^{-1}$)
    & 992 & 1527
    & Identified oscillator frequency \\

$\zeta$
    & $-0.031$ & $0.001$
    & Near-marginal oscillator damping \\

\multicolumn{4}{c}{\textit{Shear-layer forcing coefficients}} \\

$\alpha_f$
    & $-2.30\times10^4$ & $-1.90\times10^3$
    & Bulk cavity pressure forcing \\

$\alpha_r$
    & $1.11\times10^5$ & $6.10\times10^3$
    & Ramp/shock-foot forcing \\

$|\alpha_r/\alpha_f|$
    & 4.8 & 3.2
    & Relative ramp-pressure forcing strength \\

\multicolumn{4}{c}{\textit{Pressure-state dynamics}} \\

$a_f$ (s$^{-1}$)
    & $\approx 0$ & $\approx 0$
    & No independent floor-pressure relaxation \\

$a_r$ (s$^{-1}$)
    & $-146$ & $-1140$
    & Ramp-pressure relaxation rate \\

$\tau_r=1/|a_r|$ (ms)
    & 6.9 & 0.88
    & Ramp response time \\

$\Pi_r=\tau_r/T_b$
    & 1.08 & 0.22
    & Response relative to breathing period \\

$b_f$ (s$^{-1}$)
    & $8.25\times10^2$ & $6.47\times10^2$
    & Ramp $\rightarrow$ floor coupling \\

$b_r$ (s$^{-1}$)
    & $-1.60\times10^2$ & $-3.68\times10^3$
    & Floor $\rightarrow$ ramp coupling \\

\multicolumn{4}{c}{\textit{Shear-pressure coupling}} \\

$\beta_{\eta f}$ (s$^{-1}$)
    & $-3.78\times10^2$ & $1.80\times10^2$
    & Shear $\rightarrow$ floor pressure \\

$\beta_{\eta r}$ (s$^{-1}$)
    & $4.15\times10^2$ & $-1.90\times10^1$
    & Shear $\rightarrow$ ramp pressure \\

Phase ($p_r/p_f$)
    & $-133^\circ$ & $-126^\circ$
    & Measured phase lag \\

Dominant eigenmode
    & Weakly damped & Weakly damped
    & Global breathing mode \\

Secondary eigenmode
    & Strongly damped & Strongly damped
    & Pressure-subsystem dynamics
\end{tabular}

\vspace{3pt}

\caption{Comparison of identified model parameters and derived physical timescales for the low and high enthalpy flow cavity breathing modes.}
\label{tab:cold_hot_params}

\vspace{3pt}

\noindent\rule{\textwidth}{0.4pt}

\end{table*}

Although the reduced-order model is identified through linear regression, the resulting coefficients admit a direct physical interpretation in terms of the characteristic forcing and relaxation processes governing the cavity breathing dynamics. Rather than considering the individual regression coefficients in isolation, it is more instructive to group them into three physically meaningful non-dimensional parameters that characterize the coupled shear-layer--pressure feedback loop.

The first parameter compares the relaxation time of the ramp-pressure state with the characteristic breathing period,

\begin{equation}
\Pi_r=\frac{\tau_r}{T_b}
=\frac{1}{|a_r|T_b},
\label{eq:Pi_r}
\end{equation}

where $\tau_r=1/|a_r|$ is the intrinsic relaxation time of the ramp-pressure state and $T_b=1/f_b$ is the experimentally observed breathing period. This parameter measures whether the ramp-pressure field evolves quasi-steadily ($\Pi_r\ll1$), dynamically ($\Pi_r\sim1$), or slowly ($\Pi_r>1$) relative to the global breathing oscillation.

For the low enthalpy case,

\[
\Pi_r^{300}\approx1.08,
\]

indicating that the ramp-pressure state evolves on essentially the same timescale as the cavity breathing motion. Consequently, the reattachment region actively participates in the hydrodynamic feedback loop, producing the large measured phase lag between the cavity floor and ramp pressure signals.

For the high enthalpy flow,

\[
\Pi_r^{1600}\approx0.22,
\]

indicating that the ramp pressure relaxes approximately five times faster than the global oscillation period. Physically, this implies that the shock-foot and reattachment region rapidly adjust to the evolving shear layer, thereby shortening the overall feedback loop and contributing to the increased breathing frequency observed under high enthalpy-flow conditions.

The second non-dimensional parameter measures the relative forcing exerted by the two pressure states on the shear-layer oscillator,

\begin{equation}
\Gamma
=
\left|
\frac{\alpha_r}{\alpha_f}
\right|,
\label{eq:Gamma}
\end{equation}

which compares the influence of the downstream reattachment region with that of the bulk cavity pressure. For the present experiments,

\[
\Gamma^{300}\approx4.8,
\qquad
\Gamma^{1600}\approx3.2,
\]

demonstrating that the ramp-pressure fluctuations remain the dominant forcing mechanism in both operating conditions.The reduction of $\Gamma$ under high-enthalpy conditions suggests that, although the reattachment dynamics remain important, the cavity pressure field exerts a relatively stronger influence on the oscillation. The elevated temperature reduces the characteristic flow timescales and increases the speed at which pressure disturbances are communicated throughout the cavity, thereby enhancing the coupling between the cavity pressure field and the shear-layer motion.


Finally, the overall strength of the pressure-shear coupling may be characterized by

\begin{equation}
\Lambda
=
\frac{\sqrt{\beta_{\eta f}^{\,2}+\beta_{\eta r}^{\,2}}}
{\omega_0},
\label{eq:Lambda}
\end{equation}

which represents the pressure response generated by a unit shear-layer displacement normalized by the natural oscillation frequency. The parameter $\Lambda$ therefore provides a measure of the efficiency with which shear-layer motion is converted into cavity pressurization and vice versa. Although the individual coupling coefficients change with total temperature, $\Lambda$ remains of the same order of magnitude for both operating conditions, indicating that the fundamental coupling mechanism between the shear layer and the cavity pressure field is preserved.

\begin{table}
\centering
\begin{tabular*}{\columnwidth}{@{\extracolsep{\fill}}cccc}
\toprule
Parameter & Definition & $T_t \approx 300$ K & $T_t \approx 1600$ K\\
\hline

$\Pi_r$
& $\tau_r/T_b$
& 1.08
& 0.22 \\

$\Gamma$
& $|\alpha_r/\alpha_f|$
& 4.8
& 3.2 \\

$\Lambda$
& $\sqrt{\beta_{\eta f}^2+\beta_{\eta r}^2}/\omega_0$
& 0.57
& 0.12 \\

\end{tabular*}

\caption{Non-dimensional parameters extracted from the identified two-pressure-state model.}
\label{tab:nondim_params}
\end{table}


Taken together, the three parameters $(\Pi_r,\Gamma,\Lambda)$ provide a compact physical interpretation of the identified reduced-order model summarized in table~\ref{tab:nondim_params}. The cavity breathing mode is governed by (i) the relative response time of the ramp-pressure state, (ii) the competition between bulk cavity pressure and downstream shock-foot forcing, and (iii) the overall strength of the shear-layer-pressure coupling. Increasing the total temperature primarily reduces $\Pi_r$, thereby accelerating the pressure dynamics and shortening the hydrodynamic feedback loop, while leaving the overall forcing hierarchy and coupling mechanism largely unchanged.

The eigenvalue spectrum further supports this interpretation. The dominant complex-conjugate eigenpair represents the global cavity breathing mode arising from the coupled interaction between the shear layer and the cavity pressure field. The small real part of this mode indicates that the oscillator operates close to neutral stability, consistent with the experimentally observed saturated limit-cycle oscillation. The secondary eigenpair possesses substantially greater damping and corresponds to the internal adjustment of the coupled pressure subsystem. Rather than representing an independently observable oscillation, this mode governs the redistribution of pressure between the cavity interior and the reattachment region following perturbations of the shear layer. The existence of this second dynamical timescale explains the measured phase lag between the floor and ramp pressure signals and provides the physical justification for introducing two pressure states in the reduced-order model.

\section{Transient stability control via mass and heat addition}
\label{sec:control}
\subsection{Mass addition:$N_2$ Injection}

The mechanism of shear-layer oscillation shows that excess mass depletion from the cavity reduces cavity pressure. This, in turn, causes strong expansion near the cavity leading edge, resulting in the shear layer dipping into the cavity. Based on this understanding, it is hypothesized that external mass addition could influence shear-layer oscillation behaviour. 

To isolate the effect of mass addition, low enthalpy flow experiments are conducted. The combustor inlet parameter similar to low enthalpy case summarized in table~\ref{tab:test_conditions}. Once steady-state conditions are achieved, nitrogen is injected through six injector ports (three on the top wall and three on the bottom wall) located at $2H$ upstream of the cavity. The injection is performed at a momentum flux ratio of 1.85.    

\begin{figure}[!h]
\centering
\includegraphics[width=0.85\textwidth]{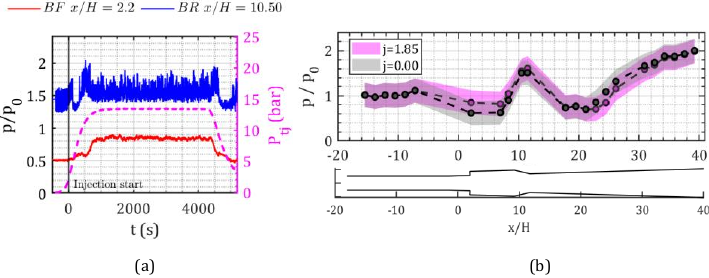}
\caption{(a) Temporal evolution of pressure inside the cavity at cavity floor (BF) and ramp (BR), along with variation of nitrogen injection pressure (b) Time-averaged bottom wall pressure variation along the combustor length.}
\label{mag_14}
\end{figure}

\begin{figure}[!h]
\centering
\includegraphics[width=0.70\textwidth]{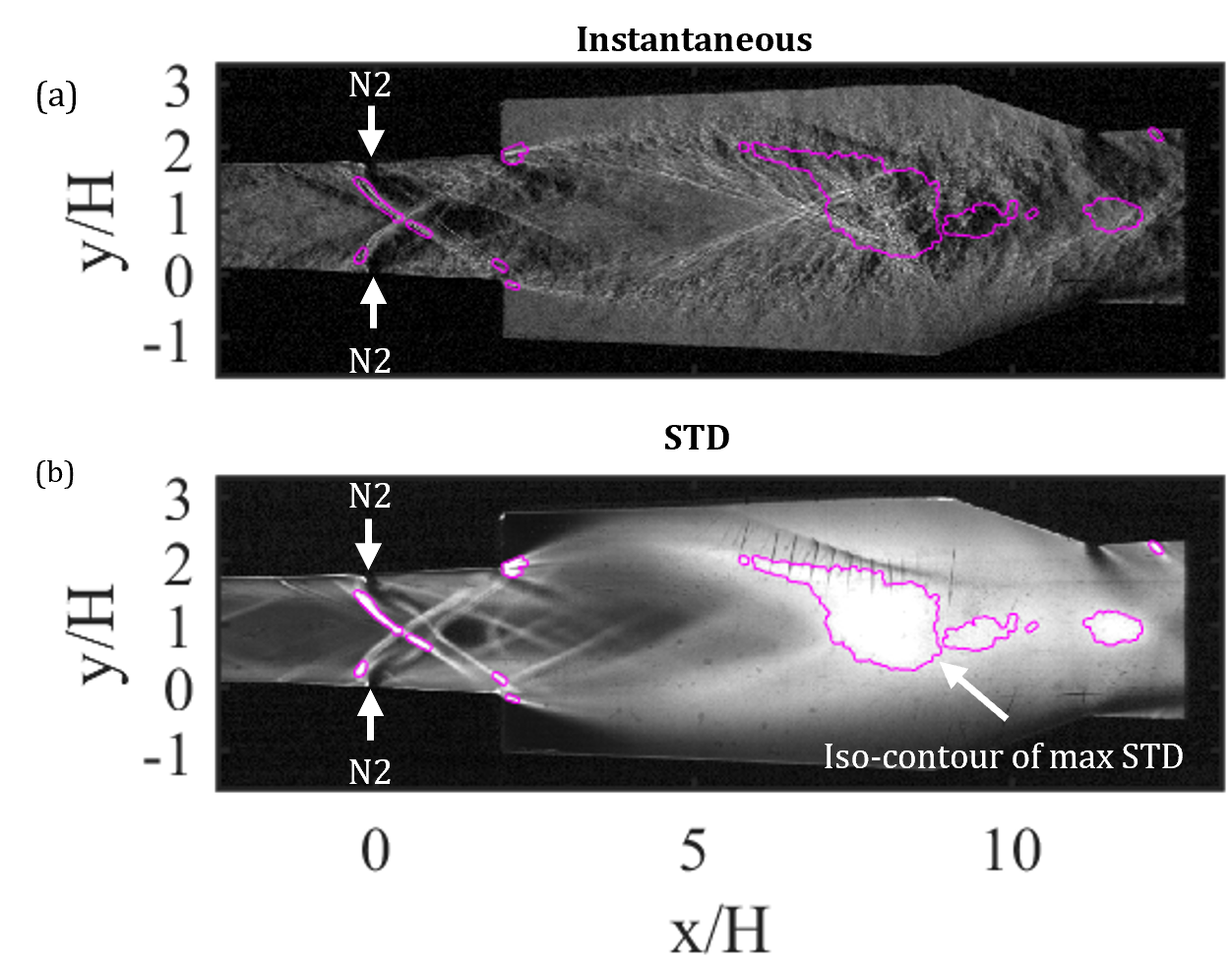}
\caption{(a) Instantaneous schileren image overlaid with iso-contour of maximum standard deviation (b) Standard deviation image highlighting $N_2$ injection in low enthalpy flow.}
\label{fig:mass_addition_schil}
\end{figure}

Figures~\ref{mag_14}(a) and (b) present the temporal evolution of the bottom cavity-floor (BF) and cavity-ramp (BR) pressures on the left axis, together with the nitrogen-injection pressure, $P_{tj}$, on the right axis. The corresponding time-averaged bottom-wall pressure distributions, with the associated standard-deviation bands, are also shown. Prior to nitrogen injection ($t < 0$ s), persistent shear-layer oscillations are observed. Upon initiation of injection, the cavity floor (BF) pressure begins to increase, temporarily perturbing the oscillation as captured by cavity ramp (BR), particularly when the injection pressure is around $5~\text{bar}$. As the injection pressure increases further and reaches steady-state conditions $\approx 15~\text{bar}$, the cavity floor (BF) pressure continues to rise up to $\approx 0.8P_{in}$ . However, the ramp pressure (BR) transitions into a different oscillatory state, and the shear-layer oscillations are not fully suppressed. 

The time-averaged pressure distribution shows a clear increase in pressure near cavity region ($0<x/H<10$), while the downstream section remains largely unaffected ($10<x/H<40$). This indicates that the influence of mass addition is predominantly localized to the cavity region, while further downstream the effect of area divergence dominates over the deceleration associated with mass addition.

The instantaneous Schlieren image and the corresponding standard-deviation field, computed over a 0.5 s interval (10000 samples), are presented in figures~\ref{fig:mass_addition_schil}(a) and (b), respectively. The instantaneous Schlieren image is processed by multiplying the measured intensity field by its streamwise density-gradient component to enhance the visibility of the shock structures. Contours corresponding to regions of high standard deviation are overlaid to identify regions of pronounced flow unsteadiness. The instantaneous image clearly captures the bow shock associated with the nitrogen injection, while the elevated standard deviation near the cavity ramp indicates persistent temporal fluctuations in this region. The persistence of enhanced fluctuations near the ramp suggests that the shear-layer dynamics continue to influence the surrounding flow field even under nitrogen injection.

Overall, the results demonstrate that while mass addition increases the cavity pressure, the magnitude of this increase is insufficient to damp the shear-layer oscillations.

\subsection{Heat addition: $C_2H_4$ Injection}
Similar to the nitrogen-injection case, high enthalpy reacting-flow experiments were conducted by replacing nitrogen with an ethylene jet and operating the preheater to establish a heated inlet condition of $T_t=1612 \pm 30$ K, a static pressure of $p=43$ kPa, and a Mach number of $M=2.47$. Once the flow conditions were established, ethylene was injected at a momentum-flux ratio of 1.38, providing sufficient jet penetration and a global equivalence ratio of $\phi_g=0.70$.

\begin{figure}[!h]
\centering
\includegraphics[width=0.85\textwidth]{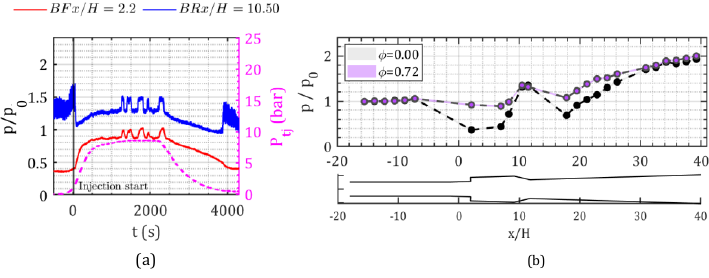}
\caption{(a) Temporal evolution of pressure inside the cavity at cavity floor and ramp, along with variation of ethylene injection pressure (b) Time-averaged bottom wall pressure variation along the combustor.}
\label{fig:heat_addition}
\end{figure}

Following injection, ethylene mixes with the crossflow and auto-ignites near the cavity ramp, after which the flame stabilizes within the cavity shear layer. Figure~\ref{fig:heat_addition}(a) and (b) presents the temporal evolution of the pressure at the cavity-floor (BF) and ramp (BR) locations, together with the ethylene injection pressure $P_{tj}$ and the time-averaged bottom-wall pressure distribution along the combustor.

Following ethylene injection, a slight increase in the cavity-floor (BF) pressure is initially observed, similar to the nitrogen mass-addition case. Once ignition occurs and the flame stabilizes within the cavity, the BF pressure increases rapidly, accompanied by a pronounced reduction in the pressure fluctuations at the cavity-ramp (BR). The attenuation of these fluctuations coincides with the suppression of the shear-layer oscillation, after which a comparatively steady shear layer is established.

\begin{figure}[!h]
\centering
\includegraphics[width=0.70\textwidth]{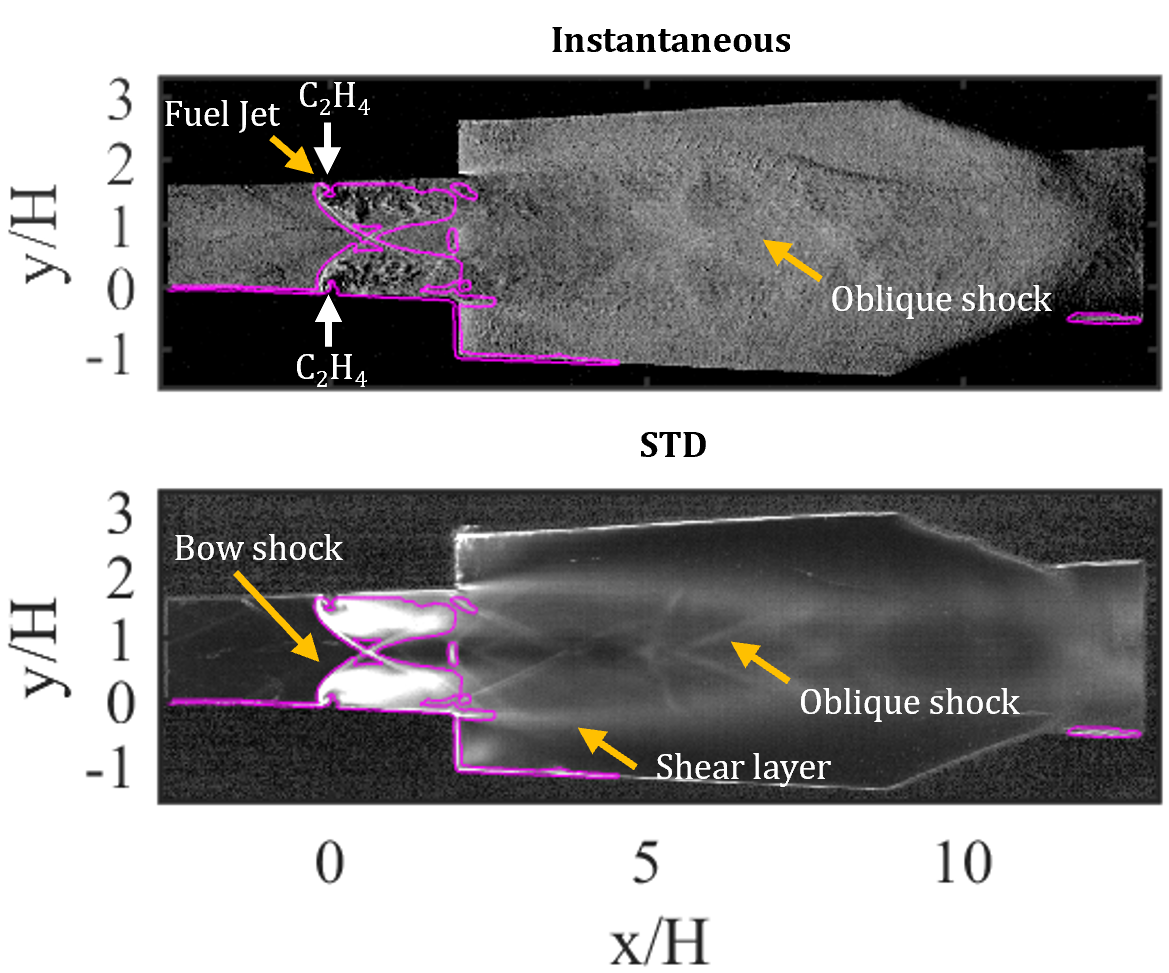}
\caption{(a) Instantaneous schileren image overlaid with iso-contour of maximum standard deviation (b) Standard deviation image highlighting $C_2H_4$ injection in high enthalpy-flow.}
\label{fig:heat_addition_schil}
\end{figure}

Figure~\ref{fig:heat_addition_schil}(a) and (b) shows the instantaneous and standard-deviation Schlieren images obtained under steady combustion conditions ($1000 < t<2500$ ms) . The standard deviation in the vicinity of the cavity ramp is substantially reduced compared with the corresponding low enthalpy condition, indicating a reduction in local flow fluctuations and providing further evidence of suppression of the shear-layer oscillation.

\begin{figure}[!h]
\centering
\includegraphics[width=0.70\textwidth]{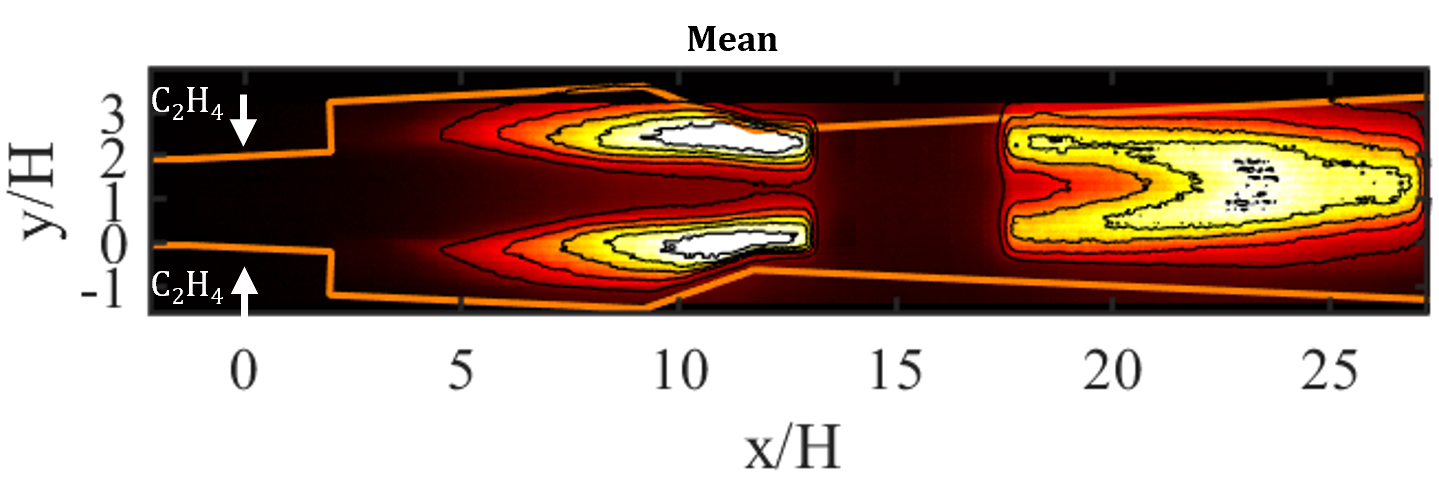}
\caption{Time-average $CH^*$ chemiluminescence image highlighting mean reaction zone.}
\label{fig:heat_addition_ch_mean}
\end{figure}

The mean $CH^*$ chemiluminescence image, shown in figure~\ref{fig:heat_addition_ch_mean}, illustrates the time-averaged flame structure. The peak $CH^{*}$ intensity is observed near the cavity ramp, followed by a long, elongated flame extending downstream, indicating sustained heat release throughout the downstream region. This spatial distribution of heat release is consistent with the measured time-averaged bottom-wall pressure distribution along the combustor.

\begin{figure}[!h]
\centering
\includegraphics[width=0.9\textwidth]{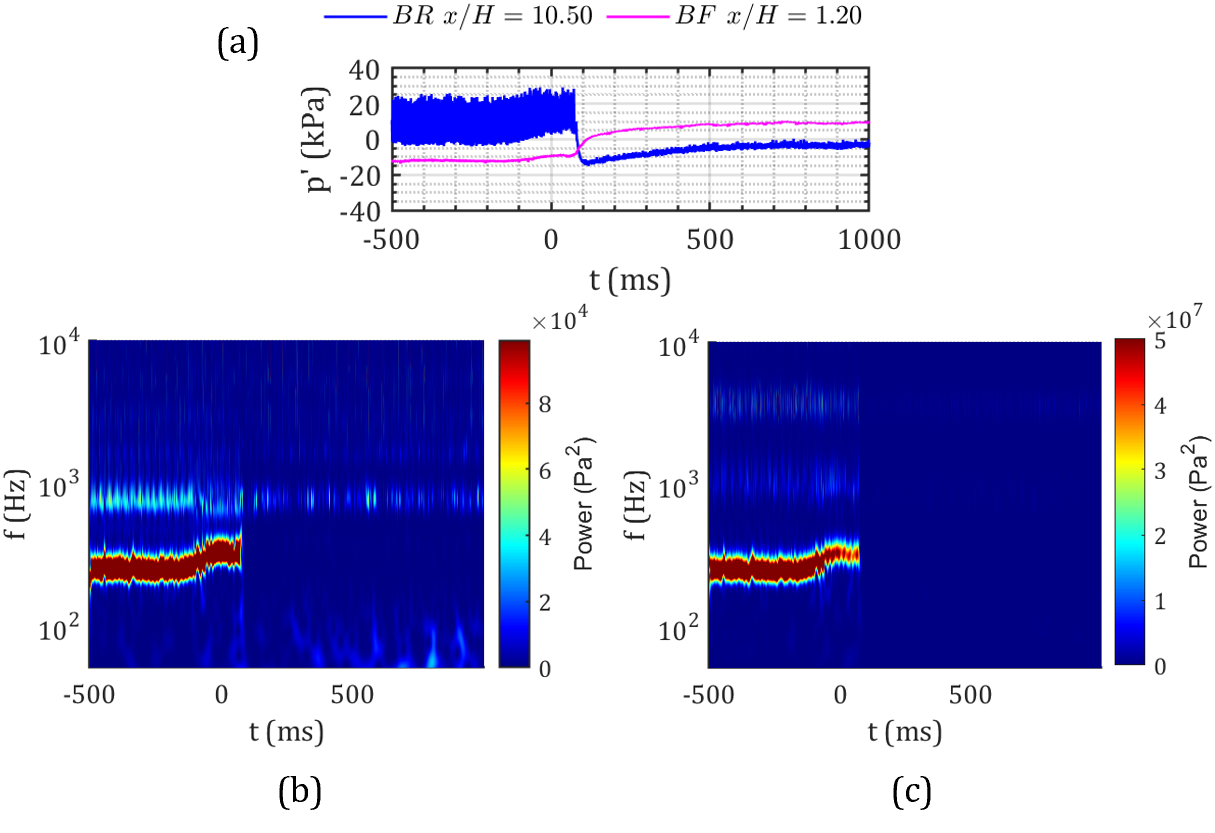}
\caption{(a) Pressure fluctuation time history at the cavity floor (BF) and ramp (BR) locations; (b) and (c) corresponding continuous wavelet transforms (CWT) of the cavity floor (BF) and ramp (BR) pressure signals.}
\label{fig:heat_addition_psd}
\end{figure}

Figure~\ref{fig:heat_addition_psd}(a) presents the pressure fluctuation ($p^{'}$) time histories at the cavity-floor (BF) and cavity-ramp (BR) locations, spanning from 500 ms before ethylene injection to 1000 ms after injection. The signals capture the transition from the pre-injection oscillatory state to the post-ignition steady state. Prior to ethylene injection, both pressure signals exhibit pronounced fluctuations associated with the shear-layer oscillation. Following ignition, the BF pressure increases rapidly, accompanied by a substantial attenuation of the pressure fluctuations at both locations. The progressive reduction in pressure fluctuations coincides with the establishment of a comparatively steady cavity pressure state and the suppression of the shear-layer oscillation.

The continuous wavelet transforms (CWTs) of the pressure signals are also shown in figures~\ref{fig:heat_addition_psd}(b) and (c). At the cavity-floor location, a pronounced frequency band is observed near the shear-layer breathing frequency ($f_b\approx244~\mathrm{Hz}$), together with intermittent higher-frequency content around $900$ Hz. A similar dominant frequency band is observed in the cavity-ramp signal prior to ignition. Following ignition, these spectral signatures are substantially attenuated, consistent with the suppression of the shear-layer oscillation. Under steady reacting conditions, the pressure signals exhibit only weak residual fluctuations, further supporting the establishment of a comparatively steady combustion state.

These results confirm that an increase in cavity pressure alone is insufficient to suppress shear-layer oscillations. Instead, heat release plays a critical role through temperature-induced dilatation, significantly reducing the pressure gradient across the shear layer and leading to complete suppression of the oscillatory behaviour.

\section{Quantitative stability analysis and growth rate extraction}

The present analysis investigates the transient modification of the cavity breathing mode following upstream injection. Two operating conditions are considered: inert $\mathrm{N_2}$ injection in the low enthalpy flow and fuel injection with subsequent heat release in the high enthalpy flow. The objective is to distinguish between transient perturbation of the cavity oscillator and permanent suppression of the breathing mode.

The cavity breathing oscillation results from the coupled interaction between the free shear layer, the cavity recirculation zone and the cavity pressure field. Pressure measurements obtained simultaneously at the cavity floor (BF) and the ramp (BR) region represent two dynamically distinct locations within the cavity. Their pressure difference,

\begin{equation}
\Delta p(t)=p_r(t)-p_f(t),
\end{equation}

provides a direct measure of the instantaneous pressure imbalance driving the shear-layer motion. Variations in $\Delta p$ therefore characterize the evolution of the cavity forcing associated with shear-layer deflection, cavity entrainment and the global breathing oscillation. Throughout this section, $\Delta p(t)$ is adopted as the principal observable for quantifying the transient cavity dynamics.

\begin{figure}[!htbp]
\centering

\begin{subfigure}[b]{0.45\textwidth}
    \centering
    \includegraphics[width=\textwidth]{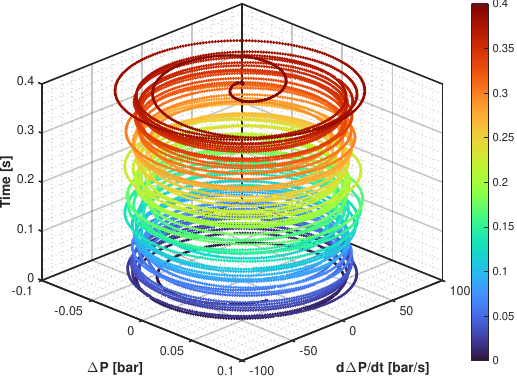}
    \caption{Limit Cycle Oscillation}
    \label{fig:hilbert}
\end{subfigure}
\hfill
\begin{subfigure}[b]{0.45\textwidth}
    \centering
    \includegraphics[width=\textwidth]{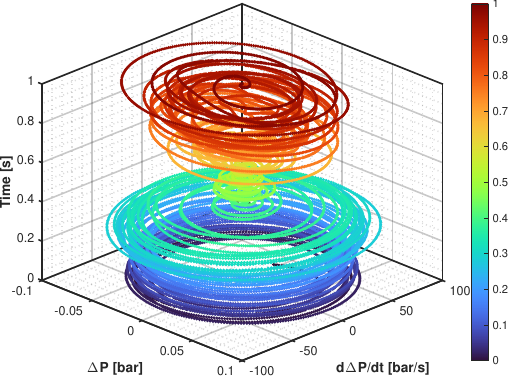}
    \caption{Temporary damping due to $N_2$ injection}
    \label{fig:phase}
\end{subfigure}
\hfill
\begin{subfigure}[b]{0.45\textwidth}
    \centering
    \includegraphics[width=\textwidth]{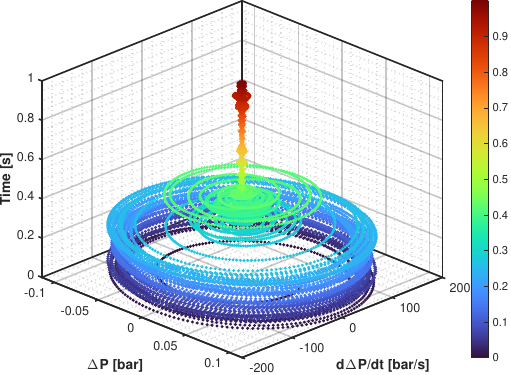}
    \caption{Permanent suppression}
    \label{fig:stability}
\end{subfigure}

\caption{Time-coloured phase-space trajectories for (a) the low-enthalpy condition, exhibiting a limit-cycle oscillation (LCO); (b) low-enthalpy flow with nitrogen injection, showing temporary attenuation of the oscillation following mass addition; and (c) high-enthalpy reacting flow with ethylene injection, showing complete decay of the oscillation following ignition and heat release.}
\label{fig:Phase-Space}
\end{figure}

Figure~\ref{fig:Phase-Space} presents the time-coloured phase-space trajectories reconstructed using $\Delta p$ and its temporal derivative, illustrating the nonlinear evolution of the cavity oscillator under the three representative operating conditions. The colour scale denotes the temporal progression of the trajectory from the initial (blue) to the final (red) state.

Low enthalpy no-injection case figure~\ref{fig:Phase-Space}(a) exhibits a nearly closed orbit of constant radius, characteristic of a saturated limit-cycle oscillation. The invariant trajectory indicates that the energy supplied through the hydrodynamic feedback loop is balanced by nonlinear dissipation, resulting in a statistically stationary oscillation with negligible net growth.

A markedly different response is observed during inert $\mathrm{N_2}$ injection low enthalpy $N_2$ injection case, figure~\ref{fig:Phase-Space}(b). Immediately after injection, the trajectory contracts towards the origin, indicating a temporary reduction in oscillation amplitude caused by mass and momentum addition to the shear layer. As the injected gas convects downstream, the perturbation weakens and the trajectory gradually expands to recover the original attractor. The restoration of the limit cycle demonstrates that inert injection perturbs the cavity oscillator without altering its underlying stability characteristics or the hydrodynamic feedback mechanism responsible for sustaining the oscillation.

In contrast, fuel injection with subsequent heat release (High enthalpy $C_2H_4$ injection case, figure~\ref{fig:Phase-Space}(c) produces a qualitatively different dynamical response. Following an initial contraction similar to the inert injection case, the trajectory continues to spiral towards the origin and ultimately collapses to a stable fixed point. The absence of any recovery indicates that heat release fundamentally modifies the coupled shear-layer--pressure feedback mechanism, driving the cavity oscillator from a self-sustained limit cycle to a stable equilibrium. Unlike the reversible perturbation produced by inert injection, combustion-induced heat release results in a permanent suppression of the breathing mode.

The contrasting attractor topologies observed in figure~\ref{fig:Phase-Space} are consistent with the reduced-order model developed in the following section. For the low enthalpy case, the identified pressure--shear coupling remains sufficiently strong ($\Lambda=0.57$) to sustain a robust limit-cycle oscillation, such that inert mass addition acts only as a transient disturbance. In the high enthalpy flow, the elevated temperature shortens the pressure-response timescale ($\Pi_{FB}=0.22$) and reduces the normalized shear--pressure coupling ($\Lambda=0.12$), leaving the system only marginally capable of sustaining the breathing mode prior to ignition. Subsequent heat release further weakens the hydrodynamic feedback loop, causing the dominant mode to become stable and the limit-cycle attractor to collapse. The quantitative relationship between these identified parameters and the cavity stability characteristics is discussed in the following sections.

\subsection{Hilbert-Transform-based amplitude extraction}

While the phase-space trajectories provide a qualitative description of the cavity attractor, they do not quantify its instantaneous stability. To characterize the temporal evolution of the breathing mode, the pressure-difference signal is analysed using the Hilbert transform, which yields the instantaneous oscillation amplitude and its associated growth rate. These quantities distinguish transient perturbations from sustained stabilization of the cavity oscillator.

The measured pressure signal contains both the coherent cavity breathing mode and broadband turbulent fluctuations. To isolate the dominant oscillation, the pressure difference $\Delta p(t)$ is band-pass filtered around the breathing frequency to obtain $\Delta p_{bp}(t)$. The corresponding analytic signal is

\begin{equation}
z(t)=\Delta p_{bp}(t)
+i\,\mathcal{H}\{\Delta p_{bp}(t)\},
\end{equation}

where $\mathcal{H}\{\cdot\}$ denotes the Hilbert transform. The instantaneous oscillation amplitude is then given by

\begin{equation}
A(t)=|z(t)|,
\end{equation}

which represents the envelope of the coherent cavity oscillation. The local stability of the breathing mode is quantified through the logarithmic growth rate,

\begin{equation}
\sigma(t)=\frac{d}{dt}\ln A(t),
\end{equation}

which follows directly from the local amplitude equation
$\dot{A}=\sigma A$. Positive, negative and near-zero values of $\sigma$ correspond to local growth, decay and saturated oscillation, respectively.

To reduce noise amplification associated with numerical differentiation, the Hilbert envelope is smoothed prior to growth-rate estimation, and the instantaneous growth rate, $\sigma(t)$, is computed using a local exponential fit over short time windows. For each analysis window, both the mean and median growth rates are computed. The median is adopted as the primary stability metric since it is less sensitive to intermittent bursts and measurement noise, whereas the mean provides a measure of the net influence of transient growth and decay events.

\begin{figure}[!htbp]
\centering
\includegraphics[width=0.9\textwidth]{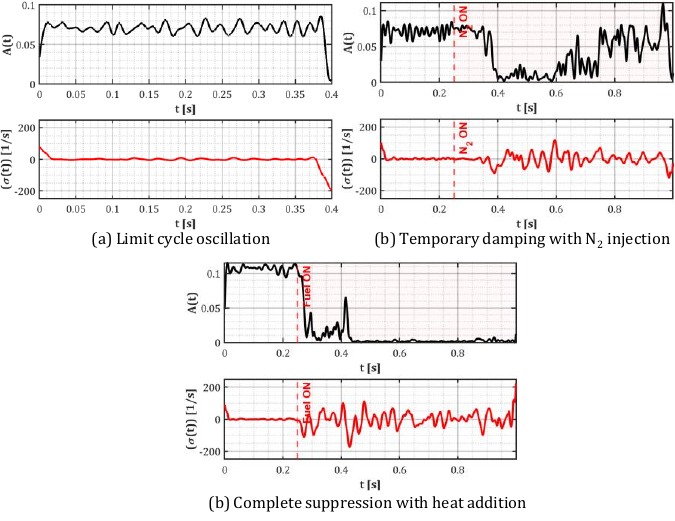}
\caption{Hilbert amplitude and instantaneous growth rate showing the temporal evolution of (a) the low-enthalpy condition exhibiting a limit-cycle oscillation (LCO); (b) low-enthalpy flow with $N_2$ addition, showing temporary attenuation of the oscillation following mass addition; and (c) high-enthalpy reacting flow with $C_2H_4$ addition, showing complete decay of the oscillation following ignition and heat release.}
\label{fig:stability_analysis}
\end{figure}

Figure~\ref{fig:stability_analysis} shows the temporal evolution of the Hilbert amplitude and the corresponding instantaneous growth rate for the three representative operating conditions.

For the low enthalpy case, the Hilbert amplitude remains nearly constant throughout the measurement period, while the growth rate fluctuates about zero with a negligible mean value. These characteristics are consistent with a statistically stationary limit-cycle oscillation in which the energy supplied by the hydrodynamic feedback loop is balanced by nonlinear dissipation.

A markedly different response is observed during $\mathrm{N_2}$ injection (low enthalpy $\mathrm{N_2}$ injection case ). Following the onset of injection, the Hilbert amplitude decreases rapidly and the growth rate becomes strongly negative, indicating temporary damping of the breathing mode caused by mass and momentum addition to the shear layer. As the injected gas convects downstream, the oscillation gradually recovers, with the amplitude returning to its pre-injection level and the growth rate relaxing towards zero. The recovery of both quantities confirms that $\mathrm{N_2}$ injection perturbs the cavity oscillator without modifying its underlying stability characteristics.

Fuel injection with subsequent heat release (high enthalpy $\mathrm{C_2H_4}$ injection case) produces a fundamentally different response. Immediately after ignition, the oscillation amplitude decreases continuously towards zero, accompanied by a sustained negative growth rate. Unlike the $\mathrm{N_2}$ injection case, neither the amplitude nor the growth rate recovers during the observation period, indicating irreversible suppression of the breathing mode. The persistent negative growth rate demonstrates that heat release fundamentally alters the coupled shear-layer-pressure feedback mechanism, driving the cavity oscillator from a self-sustained limit cycle to a stable equilibrium state.

The Hilbert amplitude and growth-rate analyses therefore provide a quantitative measure of the transient stability of the cavity oscillator. Combined with the phase-space reconstruction, they demonstrate that mass addition ($\mathrm{N_2}$ injection) produces only a reversible perturbation of the limit cycle, whereas combustion-induced heat release permanently modifies the hydrodynamic feedback loop and suppresses the breathing oscillation.

\subsection{Comparison of oscillatory states}

The Hilbert amplitude and growth-rate analyses provide an experimental measure of the stability of the cavity breathing mode and may be interpreted within the framework of the two-pressure reduced-order model. The model describes the coupled evolution of the shear-layer displacement and the cavity pressure field through the state vector

\begin{equation}
\mathbf{x}
=
\begin{bmatrix}
\eta &
\dot{\eta} &
P_c &
P_r
\end{bmatrix}^{T},
\end{equation}

where $\eta$ denotes the shear-layer displacement, while $P_c$ and $P_r$ represent the cavity floor and ramp pressure states, respectively. Linearization about the mean operating condition yields

\begin{equation}
\dot{\mathbf{x}}
=
\mathbf{A}\mathbf{x},
\end{equation}

whose eigenvalues,

\begin{equation}
\lambda_j=\sigma_j+i\omega_j,
\end{equation}

govern the modal stability and oscillation frequency. Since the Hilbert envelope satisfies the local amplitude equation,

\[
\dot{A}=\sigma A,
\]

the experimentally measured growth rate provides a direct estimate of the real part of the dominant eigenvalue,

\begin{equation}
\sigma(t)\approx
\Re\{\lambda_{\mathrm{dom}}\},
\end{equation}

thereby establishing a direct link between the measured pressure signal and the linear stability of the cavity breathing mode.

Table~\ref{tab:growth_comparison} summarizes the Hilbert amplitude and growth-rate statistics for the representative dynamical states identified in the experiments.For clarity, the distinct dynamical states are denoted as Cases A--E based on the characteristic features observed in the instantaneous growth-rate evolution, as discussed in the preceding sections.

\begin{table}
\centering
\footnotesize
\renewcommand{\arraystretch}{1.0}

\begin{tabular*}{\columnwidth}{@{\extracolsep{\fill}}
    p{0.45cm}
    p{2.3cm}
    p{1.35cm}
    p{1.35cm}
    p{1.35cm}
    p{2.55cm}
}
\toprule

Case &
State &
$\bar{A}$ &
$\bar{\sigma}$ 
($\mathrm{s^{-1}}$) &
$\tilde{\sigma}$ ($\mathrm{s^{-1}}$) &
Interpretation \\
\midrule
\hline
A &
low enthalpy flow: Oscillatory &
$6.99\times10^{-2}$ &
$-0.51$ &
$-0.89$ &
Saturated limit-cycle oscillation
\\

B &
low enthalpy flow: Peak decay &
$2.02\times10^{-2}$ &
$-13.2$ &
$-3.59$ &
Temporary suppression of the oscillatory mode
\\

C &
low enthalpy flow: Recovery &
$3.08\times10^{-2}$ &
$+9.83$ &
$+3.47$ &
Positive growth indicates re-establishment of the preferred oscillatory state.
\\

\midrule

D &
high enthalpy flow: Oscillatory &
$1.09\times10^{-1}$ &
$+0.03$ &
$+0.23$ &
Saturated limit-cycle oscillation; modal growth approximately neutral.
\\

E &
high enthalpy flow: Reacting decay &
$2.46\times10^{-2}$ &
$-19.8$ &
$-18.5$ &
Rapid suppression following ignition; strong reduction of shear-layer forcing.
\\

\bottomrule
\end{tabular*}

\caption{Comparison of cavity oscillation states using Hilbert amplitude and growth-rate statistics.}
\label{tab:growth_comparison}

\end{table}

The baseline low enthalpy and high enthalpy no injection operating conditions (Cases A and D) exhibit nearly constant oscillation amplitudes and growth rates close to zero, confirming that both flows operate as saturated limit-cycle oscillators. Although the oscillation amplitudes differ because of the different thermodynamic conditions, the near-zero median growth rates indicate that the dominant breathing mode is in a statistically stationary state, consistent with the dominant eigenvalue lying close to the imaginary axis.

The response to $\mathrm{N_2}$ injection (Cases B and C) is characterized by two distinct phases. During injection, the growth rate becomes negative, indicating temporary damping of the breathing mode caused by mass addition and pressure redistribution within the cavity. Following the passage of the injected gas, the growth rate becomes positive and the oscillation amplitude recovers towards its original value. This recovery demonstrates that $\mathrm{N_2}$ injection perturbs the cavity oscillator without altering its underlying stability characteristics. In terms of the reduced-order model, the dominant eigenvalue is displaced temporarily into the stable half-plane before returning to its near-neutral limit-cycle state.

Fuel injection ($C_2H_4$) with subsequent heat release (Case E) produces a fundamentally different response. Ignition generates a rapid reorganization of the cavity pressure field together with substantial changes in the local thermodynamic properties, including density, speed of sound and transport properties. These changes modify the coupled shear-layer-pressure feedback mechanism responsible for sustaining the breathing oscillation. Consequently, both the Hilbert amplitude and the modal growth rate decrease rapidly, with the growth rate remaining strongly negative throughout the reacting period. Unlike the inert injection case, no recovery of the oscillation is observed, indicating that heat release drives the dominant eigenmode to a stable state and permanently suppresses the cavity breathing mode.

These observations identify two fundamentally different pathways for modifying cavity oscillations,

\begin{equation}
\mathrm{Inert\ injection:}\qquad
\mathrm{LCO}
\rightarrow
\mathrm{Transient\ damping}
\rightarrow
\mathrm{Recovery},
\end{equation}

and

\begin{equation}
\mathrm{Reacting\ injection:}\qquad
\mathrm{LCO}
\rightarrow
\mathrm{Rapid\ damping}
\rightarrow
\mathrm{Stable\ equilibrium}.
\end{equation}

The distinction between these two pathways highlights the fundamentally different roles of mass addition and heat release. Whereas mass addition merely perturbs the cavity oscillator, combustion-induced heat release modifies the coupled shear-layer--pressure feedback mechanism, resulting in a permanent change in the stability characteristics of the cavity flow. The close agreement between the Hilbert growth-rate analysis and the reduced-order model demonstrates that the pressure-based stability metrics provide a quantitative experimental estimate of the dominant eigenvalue governing the cavity breathing mode.

\subsection{Hydrodynamic stability and attractor evolution}

The preceding analyses provide complementary descriptions of the cavity dynamics. The phase-space trajectories characterize the evolution of the nonlinear attractor, the Hilbert transform quantifies the instantaneous modal growth rate, and the reduced-order model identifies the corresponding dominant eigenvalue governing the cavity breathing mode. In this section, these observations are interpreted using the local hydrodynamic stability of the compressible cavity shear layer to establish a physical connection between the experimentally observed attractor evolution and the underlying flow physics.

The amplification of disturbances along the cavity shear layer is governed by the local density stratification, compressibility and pressure gradient. For a compressible, non-isothermal shear layer, the inviscid disturbance field satisfies the compressible Rayleigh equation,

\begin{equation}
    \frac{d}{dy}
    \left[
    \frac{\bar{\rho}(y)}
         {1-M_c^2(y)}
    \frac{d\hat{v}}{dy}
    \right]
    -
    k^2\bar{\rho}(y)\hat{v}(y)
    =0,
\end{equation}

where $\bar{\rho}(y)$ is the mean density, $\hat{v}(y)$ is the disturbance amplitude, $k$ is the streamwise wavenumber, and $M_c$ is the local convective Mach number. As shown in Appendix~\ref{app:rayleigh_derivation}, asymptotic integration of this equation across the thin shear layer yields the approximate scaling

\begin{equation}
\sigma
\sim
k
\left(
\frac{\bar{\rho}_{\infty}}
     {\bar{\rho}_c}
\right)
\frac{1}
{\sqrt{1-M_c^2}}
\,
\mathcal{F}(p_r-p_f),
\label{eq:growth_scaling}
\end{equation}

which relates the local hydrodynamic growth rate to the cavity density ratio, compressibility and the pressure imbalance across the cavity. Although Eq.~(\ref{eq:growth_scaling}) is obtained from a local stability analysis, it provides a physical interpretation of the experimentally measured modal growth rate and the dominant eigenvalue identified from the reduced-order model.

For the low enthalpy case, the cavity density remains comparable to the freestream density and the convective Mach number remains well below the compressibility threshold. Consequently, disturbances within the shear layer undergo sustained spatial amplification, producing a positive local growth rate that is subsequently limited by nonlinear saturation. Experimentally, this behaviour  appears as a robust limit-cycle attractor in the reconstructed phase space, a Hilbert growth rate fluctuating about zero, and a dominant eigenvalue lying close to the imaginary axis. The relatively large normalized shear--pressure coupling parameter ($\Lambda=0.57$) is consistent with strong hydrodynamic coupling between the shear layer and the cavity pressure field, allowing the oscillator to recover rapidly following the transient perturbation produced by inert $\mathrm{N_2}$ injection.

The reacting case exhibits a fundamentally different stability mechanism. Heat release substantially reduces the cavity density while simultaneously increasing the local temperature and convective Mach number. These changes modify both the compressibility and the pressure response of the cavity, reducing the hydrodynamic amplification predicted by Eq.~(\ref{eq:growth_scaling}). The reduced amplification is reflected experimentally by the continuous decrease of the Hilbert amplitude, the persistently negative modal growth rate, and the collapse of the phase-space trajectory towards a stable fixed point. Within the reduced-order model, these changes are accompanied by a reduction in the normalized shear--pressure coupling ($\Lambda=0.12$) and a substantially shorter pressure-response timescale ($\Pi_{FB}=0.22$), indicating that the pressure subsystem adjusts much more rapidly than the global breathing oscillation. The combined effect is a weakening of the coupled shear-layer--pressure feedback responsible for sustaining the cavity breathing mode.

Taken together, the four complementary analyses provide a consistent physical description of the transition. The local hydrodynamic stability determines the spatial amplification of disturbances within the shear layer; the reduced-order model translates this amplification into the global eigenstructure of the cavity oscillator; the Hilbert transform measures the corresponding temporal growth rate directly from the pressure signal; and the reconstructed phase-space trajectories reveal the resulting evolution of the nonlinear attractor. The excellent agreement between these independent analyses demonstrates that permanent suppression of the cavity breathing mode is not produced by mass addition alone, but by the combined influence of pressure-field reorganization, reduced density, increased compressibility and modified thermodynamic properties associated with combustion-induced heat release.

\section{Conclusion}
\label{sec:Conclusion}
The present study investigated the origin, evolution and suppression of low-frequency cavity breathing oscillations in an opposed dual-cavity scramjet combustor under low enthalpy and high enthalpy reacting operating conditions. Time-resolved wall-pressure measurements, synchronized Schlieren imaging, nonlinear phase-space reconstruction, Hilbert-transform analysis and a physics-based reduced-order model were combined to establish the hydrodynamic mechanism governing the cavity breathing mode and its modification by upstream injection.

Under low and high-enthalpy conditions, the present cavity configuration having large $L/H=8.0$ exhibits a self-sustained breathing oscillation arising from the coupled interaction between the free shear layer, the cavity recirculation zone and the cavity pressure field. Simultaneous pressure measurements at the cavity floor and ramp revealed a large phase difference between the two locations, demonstrating that the cavity pressure field cannot be represented by a single pressure state. Instead, the cavity dynamics are governed by the coupled evolution of a bulk cavity pressure and a reattachment/shock-foot pressure, which together drive the periodic displacement of the shear layer and sustain the global breathing mode.

A reduced-order model comprising two pressure states and a shear-layer displacement coordinate was developed to describe this coupled hydrodynamic system. The identified model accurately reproduces the experimentally observed breathing frequency, pressure phase relationship and dominant stability characteristics. The dominant eigenmode remains close to marginal stability under both low enthalpy- and high enthalpy-flow conditions, explaining the experimentally observed finite-amplitude limit-cycle oscillation. More importantly, the identified model parameters admit a direct physical interpretation through characteristic forcing and relaxation times, providing a quantitative description of the pressure--shear-layer feedback mechanism responsible for the cavity breathing dynamics.

The transient response to upstream injection revealed two fundamentally different stability pathways. Inert $\mathrm{N_2}$ injection perturbs the cavity oscillator through mass addition and pressure redistribution, producing a temporary reduction in oscillation amplitude and negative modal growth rates. However, once the injected gas convects downstream, the oscillation recovers and the cavity returns to its original limit-cycle attractor, indicating that the underlying hydrodynamic feedback mechanism remains intact. In contrast, fuel injection followed by ignition produces sustained suppression of the breathing mode. Heat release reorganizes the cavity pressure field while simultaneously modifying the density, compressibility and thermodynamic state of the cavity flow, thereby weakening the coupled shear-layer--pressure feedback responsible for sustaining the oscillation. Consequently, the Hilbert growth rate remains persistently negative and the phase-space trajectory collapses from a finite-amplitude limit cycle to a stable equilibrium.

The complementary phase-space, Hilbert-transform and reduced-order analyses provide a unified interpretation of the cavity dynamics. The reconstructed phase-space trajectories characterize the evolution of the nonlinear attractor, the Hilbert transform provides a direct experimental estimate of the modal growth rate, and the reduced-order model relates these measurements to the dominant eigenvalue governing the cavity oscillator. The local hydrodynamic stability analysis further demonstrates that the reduction in disturbance amplification under reacting conditions is consistent with the observed decrease in the normalized shear--pressure coupling parameter ($\Lambda$) and the substantially shorter pressure-response timescale ($\Pi_{FB}$), indicating that heat release fundamentally accelerates the pressure dynamics while weakening the hydrodynamic feedback loop.

The present work therefore establishes that permanent suppression of cavity breathing oscillations cannot be achieved through mass addition alone. Rather, sustained stabilization requires modification of the coupled pressure--shear-layer feedback mechanism through heat release and the accompanying changes in cavity thermodynamic properties. More broadly, the proposed framework demonstrates how experimentally measured pressure signals may be combined with nonlinear system identification and reduced-order modelling to quantify the stability characteristics of cavity oscillations directly from transient measurements. This methodology provides a physically interpretable framework for analysing unsteady combustion systems and offers a foundation for the development of model-based active control strategies for dual-mode scramjet combustors.

Overall, this work provides the first comprehensive experimental, numerical, and theoretical description of shear-layer breathing oscillations in a long, shallow opposed twin-cavity configuration. Beyond identifying the fundamental instability mechanism, it establishes a reduced-order mathematical framework closely capturing hydrodynamics time scale and demonstrates a practical suppression strategy. These findings advance the understanding of cavity-flow dynamics in supersonic combustors and provide important design guidance for stable, efficient hydrocarbon-fuelled scramjet combustion systems. 

\section*{Acknowledgments}
The authors acknowledge funding support from DRDO DFTM DIA-COE, IIT Bombay, under the Hypersonic Vertical for conducting this research. The author sumit lonkar acknowledges the Ministry of Human Resource Development, Government of India, for the graduate student scholarship. The authors also acknowledge the help and support of Bharath R., Virupaksha, and Nimesh Thakor during the experimental test campaign.

\section*{Declaration of Interests} The authors report no conflict of interest.

\bibliographystyle{jfm}
\bibliography{jfm}

@article{yu2001effect,
  title={Effect of flame-holding cavities on supersonic-combustion performance},
  author={Yu, Ken H and Wilson, Ken J and Schadow, Klaus C},
  journal={Journal of Propulsion and Power},
  volume={17},
  number={6},
  pages={1287--1295},
  year={2001}
}

@article{zhang2022experimental,
  title={Experimental study of hysteresis and catastrophe in a cavity-based scramjet combustor},
  author={Zhang, Xu and Zhang, Qifan and others},
  journal={Chinese Journal of Aeronautics},
  volume={35},
  number={10},
  pages={118--133},
  year={2022},
  publisher={Elsevier}
}

@article{vikramaditya2009effect,
  title={Effect of aft wall slope on cavity pressure oscillations in supersonic flows},
  author={Vikramaditya, NS and Kurian, J},
  journal={The Aeronautical Journal},
  volume={113},
  number={1143},
  pages={291--300},
  year={2009},
  publisher={Cambridge University Press}
}

@inproceedings{mathur2004investigation,
  title={Investigation of hydrocarbon fuels combustion in supersonic combustor},
  author={Mathur, A and Goldfeld, M and Mishunin, A and Starov, A},
  booktitle={40th AIAA/ASME/SAE/ASEE Joint Propulsion Conference and Exhibit},
  pages={3487},
  year={2004}
}

@techreport{gruber2008hydrocarbon,
  title={Hydrocarbon-Fueled Scramjet Combustor Flowpath Development for Mach 6-8 HIFire Flight Experiments (Preprint)}, 
  institution={AFRL},
  author={Gruber, Mark R and Jackson, Kevin and Liu, Jiwen},
  year={2008}
}

@article{jackson2015mach,
  title={Mach 6--8+ hydrocarbon-fueled scramjet flight experiment: the HIFiRE flight 2 project},
  author={Jackson, Kevin R and Gruber, Mark R and Buccellato, Salvatore},
  journal={Journal of Propulsion and Power},
  volume={31},
  number={1},
  pages={36--53},
  year={2015},
  publisher={American Institute of Aeronautics and Astronautics}
}

@article{sheng2024improving,
  title={Improving the combustion of scramjet engines with struts using grooves and bumps},
  author={Sheng, Zhi-Qiang and Zhang, Lan and Lu, Liang-Ze and Liu, Jing-Yuan and Hu, Xiao-An},
  journal={Aerospace Science and Technology},
  volume={147},
  pages={109047},
  year={2024},
  publisher={Elsevier}
}

@article{quan2023experimental,
  title={Experimental investigation on effects of herringbone riblets on shock wave/boundary layer interactions of a compression ramp at Mach 3},
  author={Quan, Pengcheng and Wang, Gang and Xu, Xiwang and Zhu, Ke and Yang, Yanguang},
  journal={Physics of Fluids},
  volume={35},
  number={6},
  year={2023},
  publisher={AIP Publishing}
}

@article{liu2020review,
  title={Review of combustion stabilization for hypersonic airbreathing propulsion},
  author={Liu, Qili and Baccarella, Damiano and Lee, Tonghun},
  journal={Progress in Aerospace Sciences},
  volume={119},
  pages={100636},
  year={2020},
  publisher={Elsevier}
}

@article{urzay2018supersonic,
  title={Supersonic combustion in air-breathing propulsion systems for hypersonic flight},
  author={Urzay, Javier},
  journal={Annual Review of Fluid Mechanics},
  volume={50},
  pages={593--627},
  year={2018},
  publisher={Annual Reviews}
}

@article{jazra2013design,
  title={Design of an airbreathing second stage for a rocket-scramjet-rocket launch vehicle},
  author={Jazra, Thomas and Preller, Dawid and Smart, Michael K},
  journal={Journal of Spacecraft and Rockets},
  volume={50},
  number={2},
  pages={411--422},
  year={2013},
  publisher={American Institute of Aeronautics and Astronautics}
}

@article{liu2025research,
  title={Research progress of the flow and combustion organization for the high-Mach-number scramjet: From Mach 8 to 12},
  author={Liu, Chaoyang and Ai, Junding and Zhang, Jincheng and Li, Xin and Zhao, Zijian and Huang, Wei},
  journal={Progress in Aerospace Sciences},
  volume={155},
  pages={101094},
  year={2025},
  publisher={Elsevier}
}

@inproceedings{lonkar2026experimental,
  title={Experimental Investigation of Shock Train Dynamics Under Various Fuel Injection Schemes in a Mach 2.5 Cavity Combustor},
  author={Lonkar, Sumit and Panda, Pratikash P},
  booktitle={AIAA SCITECH 2026 Forum},
  pages={0559},
  year={2026}
}

@inproceedings{thakor2020flame,
  title={Flame stabilization in high stagnation temperature supersonic flows: experiments and simulations},
  author={Thakor, Nimesh and Miranda, Cairen and Chaudhuri, Swetaprovo},
  booktitle={AIAA Scitech Forum},
  pages={1841},
  year={2020}
}

@article{bao2015effect,
  title={Effect of cavity configuration on kerosene spark ignition in a scramjet combustor at Ma 4.5 flight condition},
  author={Bao, Heng and Zhou, Jin and Pan, Yu},
  journal={Acta Astronautica},
  volume={117},
  pages={368--375},
  year={2015},
  publisher={Elsevier}
}

@article{wang2026study,
  title={Study on the cavity-based combustion of dual-mode scramjet with liquid kerosene extended at Mach 3 flight condition},
  author={Wang, Yusen and Chen, Yuqian and Tian, Ye and Huang, Yue},
  journal={Aerospace Science and Technology},
  pages={111727},
  year={2026},
  publisher={Elsevier}
}

@article{gao2024transition,
  title={Transition of the flow type in the supersonic cavity controlled by the wall temperature},
  author={Gao, Zhan and Wang, Chenglong and Sun, Yongchao and Sun, Mingbo},
  journal={International Journal of Heat and Fluid Flow},
  volume={109},
  pages={109549},
  year={2024},
  publisher={Elsevier}
}

@article{venkateswarlu2025recent,
  title={Recent advances in fuel transport and fuel-air mixing processes in supersonic combustor for scramjet applications: A review},
  author={Venkateswarlu, Kavati and Kolhe, Pankaj S and Angula, Ester},
  journal={Proceedings of the Institution of Mechanical Engineers, Part C: Journal of Mechanical Engineering Science},
  volume={239},
  number={16},
  pages={6624--6647},
  year={2025},
  publisher={SAGE Publications Sage UK: London, England}
}

@inproceedings{tuncer2010cavity,
  title={Cavity Flame Holding for High Speed Reacting Flows},
  author={Tuncer, Onur},
  booktitle={Engineering Systems Design and Analysis},
  volume={49170},
  pages={533--540},
  year={2010}
}

@article{lonkar2026mode,
  title={Mode transition and combustion-induced shock train dynamics in a cavity-based dual-mode scramjet},
  author={Lonkar, Sumit and Panda, Pratikash P},
  journal={Aerospace Science and Technology},
  pages={112194},
  year={2026},
  publisher={Elsevier}
}

@inproceedings{heller1975physical,
  title={The physical mechanism of flow-induced pressure fluctuations in cavities and concepts for their suppression},
  author={Heller, H and Bliss, D},
  booktitle={2nd Aeroacoustics conference},
  pages={491},
  year={1975}
}

@article{vishnu2019effect,
  title={Effect of heat transfer on an angled cavity placed in supersonic flow},
  author={Vishnu, AS and Aravind, GP and Deepu, M and Sadanandan, R},
  journal={International Journal of Heat and Mass Transfer},
  volume={141},
  pages={1140--1151},
  year={2019},
  publisher={Elsevier}
}

@article{rajesh2023implications,
  title={The implications of dual cavity location in a strut-mounted scramjet combustor},
  author={Rajesh, AC and Jeyakumar, S and Jayaraman, Kandasamy and Karaca, Mehmet and Athithan, A Antony},
  journal={International Communications in Heat and Mass Transfer},
  volume={145},
  pages={106855},
  year={2023},
  publisher={Elsevier}
}

@inproceedings{collatz2009dual,
  title={Dual cavity scramjet operability and performance study},
  author={Collatz, MacKenzie and Gruber, Mark and Olmstead, Dell and Branam, Richard and Lin, Kuo-Cheng and Tam, Chung-Jen},
  booktitle={45th AIAA/ASME/SAE/ASEE Joint Propulsion Conference \& Exhibit},
  pages={5030},
  year={2009}
}

@article{tan2026flame,
  title={Flame Propagation and Oscillation Characteristics in a Liquid-Kerosene-Fueled Dual-Cavity Scramjet Model Combustor},
  author={Tan, Li and Meng, Qingyang and Ren, Jie and Zhang, Zijian and Wen, Chihyung},
  journal={AIAA Journal},
  volume={64},
  number={8},
  pages={4390--4407},
  year={2026},
  publisher={American Institute of Aeronautics and Astronautics}
}

@article{wang2015large,
  title={Large eddy simulation of a hydrogen-fueled scramjet combustor with dual cavity},
  author={Wang, Hongbo and Wang, Zhenguo and Sun, Mingbo and Qin, Ning},
  journal={Acta Astronautica},
  volume={108},
  pages={119--128},
  year={2015},
  publisher={Elsevier}
}

@article{tang2024flow,
  title={Flow structures and combustion regimes in an axisymmetric scramjet combustor with high Reynolds number},
  author={Tang, Tao and Sun, Mingbo and Yan, Bo and Wang, Zhenguo and Yu, Jiangfei and Huang, Yuhui and Wang, Hongbo and Zhu, Jiajian},
  journal={Journal of Fluid Mechanics},
  volume={1000},
  pages={A3},
  year={2024},
  publisher={Cambridge University Press}
}

@article{pranaykumar2024insights,
    author = {Pranaykumar, Singeetham and Ghosh, Amardip},
    title = {Insights into flame flashback phenomenon utilizing a Strut-Cavity flame holder inside scramjet combustor},
    journal = {Physics of Fluids},
    volume = {36},
    number = {11},
    pages = {117165},
    year = {2024},
    month = {11},
    issn = {1070-6631},    
}

@article{li2020quasi,
  title={A quasi-direct numerical simulation solver for compressible reacting flows},
  author={Li, Tao and Pan, Jiaying and Kong, Fanfu and Xu, Baopeng and Wang, Xiaohan},
  journal={Computers \& Fluids},
  volume={213},
  pages={104718},
  year={2020},
  publisher={Elsevier}
}

@article{marcantoni2017rhocentralrffoam,
  title={rhocentralrffoam: An openfoam solver for high speed chemically active flows--simulation of planar detonations--},
  author={Marcantoni, LF Guti{\'e}rrez and Tamagno, Jos{\'e} and Elaskar, Sergio},
  journal={Computer Physics Communications},
  volume={219},
  pages={209--222},
  year={2017},
  publisher={Elsevier}
}

@article{zhao2016study,
  title={Study of sonic injection from circular injector into a supersonic cross-flow using large eddy simulation},
  author={Zhao, Majie and Ye, Taohong and Cao, Changmin and Zhou, Taotao and Zhu, Minming},
  journal={International journal of hydrogen energy},
  volume={41},
  number={39},
  pages={17657--17669},
  year={2016},
  publisher={Elsevier}
}

\begin{appen}
\section{Derivation of the Integrated Compressible Rayleigh Equation} 
\label{app:rayleigh_derivation}

To establish a mathematically rigorous link between the continuous fluid mechanics of the supersonic shear layer and the discrete experimental metrics (pressures $p_f, p_r$ and tracking displacement $\eta$), we integrate the inviscid, compressible Rayleigh equation across the cavity mouth interface. 

We model the shear layer as an infinitesimally thin vortex sheet localized at $y = 0$, dividing two uniform parallel streams:
\begin{align}
    \text{Freestream } (y > 0): \quad & \bar{U}(y) = \bar{U}_\infty, \quad \bar{\rho}(y) = \bar{\rho}_\infty, \quad \bar{a}(y) = \bar{a}_\infty \\
    \text{Cavity Interior } (y < 0): \quad & \bar{U}(y) = \bar{U}_c \approx 0, \quad \bar{\rho}(y) = \bar{\rho}_c, \quad \bar{a}(y) = \bar{a}_c
\end{align}
The continuous governing equation for inviscid, parallel, non-isothermal perturbations is given by:
\begin{equation}
    \frac{d}{dy} \left[ \frac{\bar{\rho}(y)}{\kappa(y)} \frac{d\hat{v}}{dy} \right] - k^2 \bar{\rho}(y) \hat{v}(y) = 0
\end{equation}
where $\hat{v}(y)$ is the complex vertical velocity perturbation amplitude, $k$ is the streamwise wavenumber, and the compressibility metric $\kappa(y)$ is defined as:
\begin{equation}
    \kappa(y) = 1 - M_\infty^2 \left( \bar{U}(y) - \frac{\omega}{k} \right)^2 \left(\frac{\bar{a}_\infty}{\bar{a}(y)}\right)^2 = 1 - M_c^2(y)
\end{equation}
Here, $\omega = \omega_r + i\sigma$ represents the complex frequency, where $\sigma$ is the growth rate extracted experimentally via the Hilbert transform.

\subsection{Control Volume Integration and Jump Conditions}
We integrate the continuous differential equation across a narrow control volume bounding the interface from $y = -\epsilon$ to $y = +\epsilon$ in the limit as $\epsilon \to 0$:
\begin{equation}
    \int_{-\epsilon}^{+\epsilon} \frac{d}{dy} \left[ \frac{\bar{\rho}(y)}{1 - M_c^2(y)} \frac{d\hat{v}}{dy} \right] dy - k^2 \int_{-\epsilon}^{+\epsilon} \bar{\rho}(y) \hat{v}(y) \, dy = 0
\end{equation}
Because the mean density $\bar{\rho}(y)$ and the vertical velocity perturbation $\hat{v}(y)$ remain bounded across the shear boundary, the second integral vanishes identically as $\epsilon \to 0$. This reduces the continuous system to a macroscopic dynamic jump condition:
\begin{equation}
    \left[ \frac{\bar{\rho}_\infty}{1 - M_{c,\infty}^2} \left(\frac{d\hat{v}}{dy}\right)_+ \right] - \left[ \frac{\bar{\rho}_c}{1 - M_{c,c}^2} \left(\frac{d\hat{v}}{dy}\right)_- \right] = 0
    \label{eq:jump_cond}
\end{equation}

\subsection{Kinematic Condition and Modal Solutions}
The matching variables at the interface must conform to the linearized kinematic boundary condition, mapping the vertical velocity directly to the macroscopically tracked shear-layer edge displacement amplitude $\eta$:
\begin{align}
    \hat{v}_+ &= i(\omega - k\bar{U}_\infty)\eta \quad (y \to 0^+) \\
    \hat{v}_- &= i\omega\eta \quad (y \to 0^-)
\end{align}
In the uniform outer zones ($y > 0$ and $y < 0$), the governing equation simplifies to a constant-coefficient wave equation, $\frac{d^2\hat{v}}{dy^2} - k^2(1 - M_c^2)\hat{v} = 0$. Enforcing bounded, decaying solutions away from the shear zone ($y \to \pm\infty$) yields:

\begin{align}
    \hat{v}(y) &= \hat{v}_+ e^{-k\sqrt{1-M_{c,\infty}^2}y} \quad \text{for } y > 0 \\
    \hat{v}(y) &= \hat{v}_- e^{+k\sqrt{1-M_{c,c}^2}y} \quad \text{for } y < 0
\end{align}
Differentiating these spatial modes with respect to $y$ provides the critical velocity gradients evaluated exactly at the boundaries of the interface:
\begin{align}
    \left(\frac{d\hat{v}}{dy}\right)_+ &= -k\sqrt{1 - M_{c,\infty}^2} \hat{v}_+ = -ik(\omega - k\bar{U}_\infty)\sqrt{1 - M_{c,\infty}^2}\eta \\
    \left(\frac{d\hat{v}}{dy}\right)_- &= +k\sqrt{1 - M_{c,c}^2} \hat{v}_- = +ik\omega\sqrt{1 - M_{c,c}^2}\eta
\end{align}

\subsection{Closed-Form Dispersion and Spatial Scale Mapping}
Substituting these normal derivative functions back into the dynamic jump relation [Eq.~\eqref{eq:jump_cond}] cancels the complex unit scalar $i$ and the geometric amplitude tracking parameter $\eta$, collapsing the system into the characteristic algebraic dispersion relation:
\begin{equation}
    \frac{\bar{\rho}_\infty (\omega - k\bar{U}_\infty)}{\sqrt{1 - M_{c,\infty}^2}} + \frac{\bar{\rho}_c \omega}{\sqrt{1 - M_{c,c}^2}} = 0
    \label{eq:dispersion_final}
\end{equation}
From the linearized transverse momentum equation, the local acoustic pressure field $\hat{p}$ links to the velocity gradients through $\frac{d\hat{p}}{dy} = -i\bar{\rho}(\omega - k\bar{U})\hat{v}$. Evaluating the macroscopic pressure difference across the mouth ($\Delta \hat{p} = \hat{p}_+ - \hat{p}_-$) yields the spatial forcing mapping:
\begin{equation}
    \Delta \hat{p} \propto k \left[ \frac{\bar{\rho}_\infty}{\sqrt{1 - M_{c,\infty}^2}} - \frac{\bar{\rho}_c}{\sqrt{1 - M_{c,c}^2}} \right] \eta
\end{equation}
Isolating the imaginary component of the complex frequency ($\sigma = \Im\{\omega\}$) from Eq.~\eqref{eq:dispersion_final} produces the closed-form scaling expression for the transient instability growth rate:
\begin{equation}
    \sigma \approx k \cdot \left( \frac{\bar{\rho}_\infty}{\bar{\rho}_c} \right) \cdot \frac{1}{\sqrt{1 - M_c^2}} \cdot \mathcal{F}(p_r - p_f)
\end{equation}
This final integrated expression provides a robust theoretical foundation for the experimental system: it proves that the growth rate $\sigma$ drops significantly when heat release reduces the cavity density $\bar{\rho}_c$, lowering the coupling coefficient $\Lambda$ to $0.12$.

\end{appen}

\clearpage

\end{document}